\documentclass[3p,11pt]{elsarticle}

\usepackage{float}

\usepackage[caption=false]{subfig}

\usepackage{graphicx}

\usepackage{booktabs}

\usepackage{amsmath}

\usepackage{amssymb}

\usepackage{nicefrac}

\usepackage{longtable}

\usepackage[ansinew]{inputenc}

\usepackage{import}

\usepackage{color}

\newcommand*{\figref}[1]{\textit{\figurename}~\ref{#1}}
\renewcommand*{\figurename}{Fig.}
\newcommand*{\tabref}[1]{\textit{\tablename}~\ref{#1}}
\renewcommand*{\tablename}{Tab.}
\let\originaleqref\eqref
\renewcommand{\eqref}{Eq.~\originaleqref}

\newcommand{\aba}{A\textsc{ba\-qus }}
\newcommand{\matlab}{M\textsc{atlab }}

\journal{Scientific Journal}

\begin{document}

\newbox\JIGAMMa
\newbox\JIGAMMb
\newbox\JIGAMMc
\newbox\JIGAMMd
\newbox\JIGAMMz
\newbox\LLbox
\newbox\LLboxh
\newbox\SLhilfbox
\newbox\SLubox
\newbox\SLobox
\newbox\SLergebnis
\newbox\TENbox
\newif\ifSLoben
\newif\ifSLunten
\newdimen\JIGAMMdimen
\newdimen\JIhsize\relax\JIhsize=\hsize
\newdimen\SLrandausgleich
\newdimen\SLhoehe
\newdimen\SLeffbreite
\newdimen\SLuvorschub
\newdimen\SLmvorschub
\newdimen\SLovorschub
\newdimen\SLsp
\def\rhotilzurho{{\JI\frac{\STAPEL\varrho!^\SLtilde\!}{\JI\!\varrho}}}
\def\rhozurhotil{{\JI\varrho\over\JI\STAPEL\varrho!^\SLtilde}}
\def\pkt{\cdot}
\def\ppkt{\mathbin{\mathord{\cdot}\mathord{\cdot}}}
\setbox\JIGAMMa = \hbox{$\scriptscriptstyle c$} \setbox\JIGAMMz =
\hbox{\hskip-.35pt\vrule width .25pt\hskip-.35pt
                        \vbox to1.2\ht\JIGAMMa{\vskip-.125pt
                             \hrule width1.2\ht\JIGAMMa height.25pt
                             \vfill
                             \hrule width1.2\ht\JIGAMMa height.25pt
                             \vskip-.125pt}%
                        \hskip-.125pt\vrule width .25pt\hskip-.125pt}
\def\Oldroy#1#2#3{\STAPEL{#1}!_\SLstrich!_\SLstrich!^\circ{}
                  \ifx #2,{}_{\copy\JIGAMMz}%
                  \else \mskip1mu{}^{\copy\JIGAMMz}\fi
                  \mskip1mu\ifx #3,{}_{\copy\JIGAMMz}%
                           \else {}^{\copy\JIGAMMz}\fi }
\def\OP#1#2{\ifnum#1=1{\rm S}
            \else\ifnum#1=2{\rm S}^\JIv
                 \else\ifnum#1=3{\rm S}^T
                      \else{{\rm S}^T}^\JIv
            \fi\fi\fi\LL{{#2}}\RR}
\edef\JIminus{{\setbox\JIGAMMa=\hbox{$\scriptstyle x$}%
           \hbox{\hskip .10\wd\JIGAMMa
                 \vbox{\hrule width .6\wd\JIGAMMa height .07\wd\JIGAMMa
                       \vskip.53\ht\JIGAMMa}%
                 \hskip .10\wd\JIGAMMa}}}
\edef\JIv{{\JIminus 1}}
\def\JI{\displaystyle}
\def\JIha{{1\over 2}}
\def\JIfolgt{\quad\Rightarrow\quad}
\def\LL#1\RR{\setbox\LLbox =\hbox{\mathsurround=0pt$\displaystyle
                                              \left(#1\right)$}%
       \setbox\LLboxh=\hbox{\mathsurround=0pt%
                  $\displaystyle{\left(%
                      \vrule width 0pt height\ht\LLbox depth\dp\LLbox
                      \right)}$}%
       \left(\hskip-.3\wd\LLboxh\relax\copy\LLbox
              \hskip-.3\wd\LLboxh\relax\right)}
\def\ZBOX#1#2#3{\def#3{}%
                \setbox#1 = #2
                \def#3{ to \wd#1}%
                \setbox#1 = #2}
\def\SLdreieck{\setbox\TENbox=\hbox{\fontscsy\char 52}
                 \dp\TENbox = 0pt
                 \hbox{\hskip -2\SLrandausgleich
                       \box\TENbox
                       \hskip -2\SLrandausgleich}}
\def\SLtilde{\setbox\TENbox=\hbox{\fontscex\char 101}
                    \vbox{\vskip-.03\ht\TENbox
                          \hbox{\hskip -1\SLrandausgleich
                                \copy\TENbox
                                \hskip -1\SLrandausgleich}
                          \vskip -.86\ht\TENbox}}
\def\SLstrich{\vrule width \SLeffbreite height.4pt}
\def\SLpunkt{{\vbox{\hbox{$\displaystyle.$}\vskip.03cm}}}
\def\SLabstand{\vskip .404pt}
\def\SLzwischen{\vskip 1.372pt}
\font\fontscsy=cmsy6 \font\fontscex=cmex10 scaled 1200
\def\STAPEL#1{\def\SLkern{#1}%
              \futurelet\next\SLpruef
               A_0   _0    :B_0   _-.17 :C_.05 _-.15 :D_0   _-.2
              :E_0   _-.2  :F_0   _-.21 :G_0   _-.15 :H_0   _-.23
              :I_.2  _.15  :J_.05 _-.1  :K_0   _-.22 :L_0   _-.1
              :M_0   _-.23 :N_0   _-.25 :O_.05 _-.2  :P_0   _-.21
              :Q_.05 _-.2  :R_0   _-.03 :S_0   _-.15 :T_.2  _0
              :U_.1  _-.1  :V_.1  _-.15 :W_.1  _-.2  :X_0   _-.22
              :Y_.16 _-.15 :Z_0   _-.25
              :a_.05 _-.05 :b_.05 _0    :c_.05 _.05  :d_0   _-.05
              :e_.07 _0    :f_0   _-.15 :g_.04 _-.2  :h_0   _-.07
              :i_.05 _0    :j_.08 _-.1  :k_0   _-.1  :l_.2  _.15
              :m_0   _-.1  :n_0   _-.1  :o_0   _-.1  :p_.15 _0
              :q_.1  _0    :r_.1  _-.1  :s_0   _-.2  :t_.1  _.05
              :u_0   _-.1  :v_0   _-.2  :w_0   _-.2  :x_.04 _-.14
              :y_.15 _-.05 :z_0   _-.15
							:\mathcal{K}_.05_-.24 :\psi_0.06 _-0.2
              :\mit\Phi_.08 _-.1    :\mit\Omega_0 _-.2   :\varXi_.00 _-.2
              :\alpha_0 _-.2        :\gamma_.1 _-.1      :\varepsilon_.1 _-.1
              :\epsilon_.05 _-.05   :\eta_.05 _-.15      :\lambda_0 _0
              :\mu_0 _-.25          :\nu_0 _-.2          :\varSigma_-.03 _-.2
              :\varrho_.00 _-.2     :\sigma_.1 _-.2      :\tau_.15 _-.15
              :\varphi_.2 _-.1      :\omega_.1 _-.1      :\mit\Gamma_-.1 _-.1
              :\Lambda_0 _0         :\Gam_0 _0           :\Lam_0 _0
              :\SLsuchende
              \def\SLtrick{\noexpand\SLtrick\noexpand}%
                \def\SLdummy{\noexpand\SLdummy}%
                \edef\SLoboxinhalt{}\edef\SLuboxinhalt{}%
                \SLobenfalse\SLuntenfalse
                \futurelet\next\SLsuchruf}
  \def\SLsuchruf{\ifx\next !\let\next\SLexpand
                 \else\let\next\SLerzeug\fi\next}
  \def\SLexpand#1#2#3{\ifx #2\sb\ifSLunten\let\SLspeicher\SLuboxinhalt
                   \else\def\SLspeicher{\SLtrick\SLabstand}\fi
                   \edef\SLuboxinhalt{%
                       \SLspeicher
                       \SLtrick\SLzwischen
                       \hbox\SLdummy{\hfil\mathsurround=0pt
$\SLtrick\scriptstyle\SLtrick#3$%
                                     \hfil}}%
                   \SLuntentrue%
                      \else\ifSLoben\let\SLspeicher\SLoboxinhalt
                   \else\def\SLspeicher{\SLtrick\SLabstand}\fi
                   \edef\SLoboxinhalt{%
                       \hbox\SLdummy{\hfil\mathsurround=0pt
$\SLtrick\scriptstyle\SLtrick#3$%
                                     \hfil}%
                       \SLtrick\SLzwischen
                       \SLspeicher}%
                   \SLobentrue\fi\futurelet\next\SLsuchruf}
  \def\SLerzeug{\def\SLtrick{}
                \setbox\SLhilfbox=\hbox{$\displaystyle{E}$}%
                \SLrandausgleich=.04\wd\SLhilfbox
                      \setbox\SLhilfbox=%
                         \hbox{\hskip -1\SLrandausgleich
                          \mathsurround=0pt$\displaystyle{\SLkern}$%
                               \hskip -1\SLrandausgleich}%
                      \SLhoehe = \ht\SLhilfbox
                      \advance\SLhoehe by \dp\SLhilfbox
                      \SLeffbreite = \wd\SLhilfbox
                      \advance\SLeffbreite by \SLab\SLhoehe
                      \ZBOX\SLubox{\vbox{\offinterlineskip
                                         \SLuboxinhalt
                                         \hrule height 0pt}}\SLdummy
                      \ZBOX\SLobox{\vbox{\offinterlineskip
                                         \SLoboxinhalt
                                         \hrule height 0pt}}\SLdummy
                      \SLsp = \SLzu\SLhoehe
                      \advance\SLsp by -.5\SLeffbreite
                      \SLuvorschub = -1\SLsp
                      \advance\SLuvorschub by -.5\wd\SLubox
                      \SLovorschub = -1\SLsp
                      \advance\SLovorschub by -.5\wd\SLobox
                      \advance\SLovorschub by .26\SLhoehe
                      \ifdim\SLuvorschub > \SLovorschub
                         \SLsp = \SLovorschub
                      \else
                         \SLsp = \SLuvorschub
                      \fi
                      \ifdim\SLsp < 0pt%
                         \advance\SLuvorschub by -1\SLsp
                         \SLmvorschub = -1\SLsp
                         \advance\SLovorschub by -1\SLsp
                      \else
                         \SLmvorschub = 0pt
                      \fi
                      \setbox\SLergebnis = \hbox{%
                         \offinterlineskip
                         \hskip\SLrandausgleich\relax
                         \vbox to 0pt{%
                            \vskip -1\ht\SLobox
                            \vskip -1\ht\SLhilfbox
\hbox{\hskip\SLovorschub\copy\SLobox\hfil}%
                            \hbox{\hskip\SLmvorschub\copy\SLhilfbox
                                  \hfil}%
\hbox{\hskip\SLuvorschub\copy\SLubox\hfil}%
                            \vss}%
                         \hskip\SLrandausgleich}%
                      \SLsp = \dp\SLhilfbox
                      \advance\SLsp by \ht\SLubox
                      \dp\SLergebnis = \SLsp
                      \SLsp = \ht\SLhilfbox
                      \advance\SLsp by \ht\SLobox
                      \ht\SLergebnis = \SLsp
                      \box\SLergebnis{}}
  \def\SLpruef{\ifx\next\SLsuchende\def\SLzu{0}\def\SLab{0}%
                  \def\next##1\SLsuchende{\relax}%
               \else\let\next\SLvergl
               \fi\next}
  \def\SLvergl#1_#2_#3:{\def\SLv{#1}%
                        \ifx\SLkern\SLv\def\SLzu{#2}\def\SLab{#3}%
\def\next##1\SLsuchende{\relax}%
                        \else\def\next{\futurelet\next\SLpruef}
                        \fi\next}
\def\PKT#1{#1!^\SLpunkt}

\newbox\minusbox
\def\minus{\mathchoice{\minusarb\displaystyle}%
                      {\minusarb\textstyle}%
                      {\minusarb\scriptstyle}%
                      {\minusarb\scriptscriptstyle}}
  \def\minusarb#1{\setbox\minusbox=\hbox{$#1x$}%
                  \hbox{\hskip .10\wd\minusbox
                        \vbox{\hrule width .6\wd\minusbox
                                     height .07\wd\minusbox
                              \vskip.53\ht\minusbox}%
                        \hskip .10\wd\minusbox}}

\def\Basis{\STAPEL e!_\SLstrich}
\def\C{\STAPEL C!_\SLstrich!_\SLstrich}
\def\X{\STAPEL X!_\SLstrich!_\SLstrich}
\def\Cinv{\STAPEL C!_\SLstrich!_\SLstrich^{\minus 1}}
\def\Cg{\STAPEL C!_\SLstrich!_\SLstrich!^\SLstrich}
\def\CD{\STAPEL C!_\SLstrich!_\SLstrich!^\SLdreieck}
\def\XD{\STAPEL X!_\SLstrich!_\SLstrich!^\SLdreieck}
\def\ECKMT{\STAPEL M!_\SLstrich!_\SLstrich!^\SLdreieck^T}
\def\CgD{\STAPEL C!_\SLstrich!_\SLstrich!^\SLstrich!^\SLdreieck}
\def\BD{{\STAPEL C!_\SLstrich!_\SLstrich!^\SLdreieck}{^{\JIv}}}
\def\BDn#1{{\STAPEL C!_\SLstrich!_\SLstrich!^\SLdreieck}{^{\JIv}_{#1}}}
\def\Ttil{\STAPEL T!_\SLstrich!_\SLstrich!^\SLtilde}
\def\rhotil{\JI\STAPEL\varrho!^\SLtilde\!}
\def\rhotilzurho{{\JI\frac{\STAPEL\varrho!^\SLtilde\!}{\JI\!\varrho}}}
\def\rhozurhotil{{\JI\varrho\over\JI\STAPEL\varrho!^\SLtilde}}
\def\Skal#1{\STAPEL {#1}}
\def\Vek#1{\STAPEL {#1}!_\SLstrich}
\def\Ten2#1{\STAPEL {#1}!_\SLstrich!_\SLstrich}
\def\F{\STAPEL F!_\SLstrich!_\SLstrich}
\def\Fg{\STAPEL F!_\SLstrich!_\SLstrich!^\SLstrich}
\def\e{\STAPEL e!_\SLstrich!_\SLstrich}
\def\b{\STAPEL b!_\SLstrich!_\SLstrich}
\def\D{\STAPEL D!_\SLstrich!_\SLstrich}
\def\I{\STAPEL I!_\SLstrich!_\SLstrich}
\def\I{\STAPEL I!_\SLstrich!_\SLstrich}
\def\L{\STAPEL L!_\SLstrich!_\SLstrich}
\def\CnOldWed{\overset{\circ{}}{\Big(\STAPEL{C}!_\SLstrich!_\SLstrich_2\Big)}
                  \mskip1mu\mskip1mu
                  \mskip1mu {}^{\hat{\copy\JIGAMMz}}
                  \mskip1mu {}_{\hat{\copy\JIGAMMz}}}
\def\CnOldWedred{\STAPEL{C}!_\SLstrich!_\SLstrich!^\circ{}_2}

\def\CD{\C!^\SLdreieck}
\def\CDn#1{\C!^\SLdreieck_#1}
\def\X{\Ten2 X}
\def\XDn#1{\X!^\SLdreieck_#1}
\def\Cauchy{\STAPEL \sigma!_\SLstrich!_\SLstrich}
\def\Cn#1{\C_{#1}}
\def\Ln#1{\L_{#1}}
\def\Fn#1{\F_{#1}}
\def\qtil{\STAPEL q!_\SLstrich!^\SLtilde}
\newcommand{\IntGtil}[1]
   {\displaystyle{{\Big.^{\widetilde{\mathcal{G}}}}
      \hspace{-1.5ex}\int #1 \ {\rm d}\widetilde{V}}}
\newcommand{\IntGtilN}[2]
   {\displaystyle{{\Big.^{\widetilde{\mathcal{G}}_{#1}}}
      \hspace{-1.5ex}\int #2 \ {\rm d}\widetilde{V}_{#1}}}
\newcommand{\IntGhat}[1]
   {\displaystyle{{\Big.^{\widehat{\mathcal{G}}}}
      \hspace{-1.5ex}\int #1 \ {\rm d}\widehat{V}}}
\newcommand{\nadel}[1]
   {\big(\hspace{-3pt}\big(\hspace{1pt} #1\hspace{1pt}\big) \hspace{-3pt}\big)}
  
\newcommand{\inv}{{\minus 1}}
\newcommand{\invT}{{\minus\T}}    
\def\indLT#1#2%
  {\mathop{}%
   \mathopen{\vphantom{#2}}^{\scriptscriptstyle #1}%
   \kern-\scriptspace%
   #2}
	
\def\Hav{\Ten2H!^*}	
\def\Fav{\Ten2F!^*}	
\def\Tav{\Ten2T!^*}

\newcommand{\FavKoef}[1]{\Skal{F}!^*_{#1}}
\newcommand{\TavKoef}[1]{\Skal{T}!^*_{#1}}
	
\newcommand{\Cstiff}{\STAPEL C!_\SLstrich!_\SLstrich!_\SLstrich!_\SLstrich}
\newcommand{\Mstiff}{\STAPEL M!_\SLstrich!_\SLstrich!_\SLstrich!_\SLstrich}
\newcommand{\abastiff}{\frac{1}{J_3}\STAPEL k!_\SLstrich!_\SLstrich!_\SLstrich!_\SLstrich^*}
\newcommand{\FPD}{\textit{FPD}}
\newcommand{\Kvier}{\STAPEL K!_\SLstrich!_\SLstrich!_\SLstrich!_\SLstrich}

\newcommand{\Ceins}{\Ten2C!^\SLstrich}
\newcommand{\Ein}{\Ten2I}
\newcommand{\A}{\Ten2A!^\SLtilde}
\newcommand{\Jhoch}[2]{J_3^{\nicefrac{#1}{#2}}}

\newcommand{\dya}{\otimes}
\newcommand{\simu}[1]{\textit{Sim. #1}}


\begin{frontmatter}

\title{Evaluation of hyperelastic models for unidirectional short fibre reinforced materials using a representative volume element with refined boundary conditions}

\author{N.~Goldberg}
\author{H.~Donner}
\author{J.~Ihlemann}

\address{Technische Universit\"at Chemnitz, Professorship of Solid Mechanics, Reichenhainer Str. 70, 09126 Chemnitz, Germany}

\begin{abstract}
The simulation of a short fibre reinforced structure by means of the FEM requires the knowledge of the material behaviour at every Gauss point. In order to obtain such information, a representative volume element (RVE) containing unidirectional short fibres is analysed in the presented work. The findings are used to assess the applicability of several hyperelastic models describing transversal isotropic materials under consideration of large deformations. As the RVE's average response represents the homogenised behaviour at a macroscopic material point, the material models' parameters can be identified by fitting them to stress-strain curves obtained from simulations with the RVE. The application of periodic boundary conditions to the RVE in tensorial form enables a simple access to consider tilted fibres and catch the anisotropy induced by the fibres. The comparison of the calibrated material model with the RVE allows the assessment of the material model's applicability and quality. Both the modelling and the calculations are carried out with the commercial FEM software A\textsc{ba\-qus}.
\end{abstract}

\begin{keyword}

RVE \sep short fiber reinforced structure \sep homogenization \sep unidirectional \sep anisotropy \sep parameter identification

\end{keyword}

\end{frontmatter}


\section{Introduction}
\begin{figure}[htb]
\centering
\includegraphics[width = \textwidth]{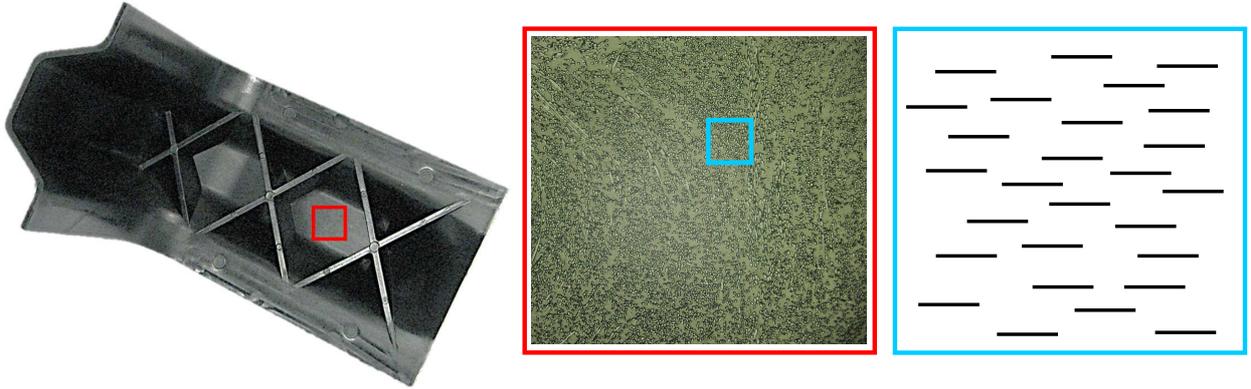}
\caption[]{Different scales of a short fibre reinforced U-bend with reinforcing ribs (left). The light microscopy image (middle) reveals groups of fibres (top view), in which the fibres are aligned unidirectional (shown in the side view on the right). Source: Federal Cluster of Excellence EXC 1075}
\label{fig:motivation}
\end{figure}

Despite showing homogeneous properties on the macroscopic scale, short fibre reinforced materials are made up of heterogeneities on a microscopic scale as it can be seen in \figref{fig:motivation}. In order to adequately simulate such a structure using the FEM, a continuum mechanical model for every Gauss point is required. Such a model should include the characteristics of the material behaviour at the macroscopic scale, meaning it has to represent the features of a matrix material with fibres. The investigation of those features can be done taking advantage of a representative volume element (RVE). 

The RVE contains a representative amount of heterogeneities (short fibres) whose local responses to external loads or displacements differ from the RVE's response. However, the averaging of these local responses over the RVE provides a homogeneous response, which equals the global characteristics of the RVE. The RVE's size plays a significant role. On the one hand it has to be larger than the heterogeneous structures it contains in order to include a sufficient amount of them. On the other hand it must be small enough to be treated as an infinite point on the macroscopic scale. The Hill condition \citep{Hill.1963} evaluates the RVE's size in such a manner. This criterion reaches back to 1963 and proposes the equivalence between the deformation energy on the macro scale and the overall deformation energy of the RVE. Later on the condition was extended to a similar equivalence between the stress power on micro and macro scale \citep{Gluge.2013}.

One of the main characteristics of the short fibre reinforced material is its anisotropy induced by the high length-to-width-ratio of the fibres. In order to cover these anisotropic features, the RVE has to be investigated under different angles between the fibres' direction and the direction of the external load and displacement. Several approaches to achieve this task can be found in the literature. Ideally, a RVE would model the micro structure in complete accordance to the real material. Due to limitations in the modelling and calculation capacities, the RVE usually contains only a discretisation of the actual topology. A well established way of doing this is the use of two- (2D) or three-dimensional (3D) finite element models. The publications \citep{Xia.2003, Chen.2004, Abadi.2012} deal with RVEs consisting of a single fibre surrounded by matrix material whereas \citep{Shan.2002, Gonzalez.2007} are focused on 2D RVEs with multiple fibres. The paper \citep{Wang.2011} deals with a voxel-based approach to model RVEs with short cylindrical inclusions randomly distributed in the 3D space. In \citep{Gluge.2012} spheric particles in cubic and spheric RVEs are considered. Another approach was advocated by \citep{Reese.2003} where fibres were discretisised by wire elements. Up to the authors' knowledge, all approaches in the literature require the modelling of actually rotated fibres in order to capture the anisotropic properties.

In the present work, a RVE, which allows the consideration of multiple fibre angles, is introduced. The RVE excels with its simpleness and flexibility. Special consideration is given to the choice of the boundary conditions and their implementation into the FEM software A\textsc{ba\-qus}. The RVE is subjected to several deformations in order to generate synthetic reference values and demonstrate the RVE's anisotropic properties. The synthetic data are taken to identify the material parameters of several hyperelastic material models. The comparison of the data obtained by the RVE and the predicted data by the material model with identified parameters is then used to assess the applicability of the constitutive assumptions of the material models.

\subsection{Notation}
The present paper makes use of a coordinate-free tensor notation. The number of bars under the tensor symbol represents the tensor's order, i.e. vectors read for example $\Vek{r}$ and second-order tensors $\Ten2X$. The tensor's coefficients are related to orthonormal base vectors, e.g. $\Ten2X = X_{ab}\Vek{e}_a \dya \Vek{e}_b$. The base vectors $\left.\Vek{e}_a\right|_{a=X,Y,Z}$ refer to the global coordinate system whereas base vectors, corresponding to the local coordinate system (see section \ref{chap:modelling}), are distinguished by a different index $\left.\Vek{e}_i\right|_{i=1,2,3}$. The treatment of the indices corresponds to the Einstein summation convention. The tensor product is symbolised by '$\dya$'. The operator '$\pkt$' describes a single contraction between base vectors. Multiple dots mean multiple contractions such that
\begin{equation}
\begin{aligned}
(\Vek{e}_a \dya \Vek{e}_b) \pkt (\Vek{e}_c \dya \Vek{e}_d) &= (\Vek{e}_b \pkt \Vek{e}_c)\Vek{e}_a \dya \Vek{e}_d = \delta_{bc}\Vek{e}_a \dya \Vek{e}_d \qquad , \\
(\Vek{e}_a \dya \Vek{e}_b) \ppkt (\Vek{e}_c \dya \Vek{e}_d) &= (\Vek{e}_b \pkt \Vek{e}_c)(\Vek{e}_a \pkt \Vek{e}_d) = \delta_{bc}\delta_{ad} \qquad ,
\end{aligned}
\end{equation}
where $\delta$ is the Kronecker delta. $\Ten2I$ is the second-order idenditiy tensor. The Frobenius norm of a tensor $\Ten2X$ is denoted by $\|\Ten2X\|$. Tensors in the reference configuration read $\Ten2X!^\SLtilde$. The material time derivative $\Ten2X!^\SLdreieck$ is defined by
\begin{equation}
\begin{aligned}
\Ten2X!^\SLdreieck = \STAPEL X!^\SLpunkt_{ab}\Vek{e}_a \dya \Vek{e}_b \qquad,
\end{aligned}
\end{equation}
with fixed base vectors. Averaged macroscale values are labelled as $\Ten2X!^*$. Furthermore, a symmetry operator for fourth-order tensors is introduced as follows
\begin{equation}
\begin{aligned}
\Kvier^\mathrm{S_{24}} &= \frac{1}{4}\left( K_{adcb} + K_{dacb} + K_{adbc} + K_{dabc} \right)\Vek{e}_a\dya\Vek{e}_b\dya\Vek{e}_c\dya\Vek{e}_d
\end{aligned}
\end{equation}
The deviator of a second-order tensor reads
\begin{equation}
\begin{aligned}
\Ten2X^D = \Ten2X - \frac{1}{3}(\Ein \ppkt \Ten2X)\Ein \qquad .
\end{aligned}
\end{equation}

\subsection{List of symbols}

\begin{longtable}{rp{.8\textwidth}}
\toprule
$\mathcal{K}, \Skal{\mathcal{K}}!^\SLtilde, \Skal{\mathcal{K}}!^\wedge$ & current, reference and intermediate configurations \\
$\Ten2F, \C$              & deformation gradient, right Cauchy-Green tensor\\
$\Vek{a}!^\SLtilde$       & fibre orientation in reference configuration \\ 
$\A$                      & structural tensor \\
$J_1,J_2,J_3$             & invariants of $\C$\\
$I_4,I_5,J_4,J_5$         & mixed invariants of $\C$ and $\A$\\
$\psi$                    & energy density \\
$\Ten2\sigma, \Ten2T, \Ten2T!^\SLtilde$	& Cauchy stress tensor, first and second Piola-Kirchhoff stress tensors \\
$\Cstiff, \Mstiff$        & Eulerian and Lagrangian stiffness tensors \\
$\abastiff$               & stiffness tensor required by \aba \\
$G,K,E_F$                 & shear modulus, bulk modulus and Young's modulus of fibre in the material model\\
$\lambda_F$               & stretch of fibre \\
$f_{\mathrm{vol}}$				& fibre volume fraction \\
$d_{\mathrm{fib}}$				& minimum distance between fibres \\
$FPD$											& average number of fibres per spatial direction \\
$\chi$										& length-to-width-ratio of single fibre \\
$l_1, l_2, l_3$						& pilot nodes' distance to local origin \\
$\Vek{u}$     						& displacement vector \\
$\Vek{x}$, $\Delta\Vek{x}$& position vector, difference of two position vectors \\
$\Hav, \Fav, \Tav$ 				& averaged values of: displacement gradient, deformation gradient and first Piola-Kirchhoff stress tensors \\
$a_1,a_2,a_{ij}$          & scalar factors for linear constraints in \aba \\
$\varphi$									& fibre angle \\
$P_s$                     & stress power \\
$v_i$											& coefficient of velocity \\
$\STAPEL V!^\SLtilde$			& reference volume \\
$n_\bot, n_\parallel$     & number of elements per fibre in transversal and longitudinal direction \\
$\lambda, \kappa$		      & stretch and shear \\
$G_M$/$G_F$								& shear modulus of matrix/fibre in the RVE\\
$K_M$/$K_F$								& bulk modulus of matrix/fibre in the RVE\\
\bottomrule
\end{longtable}


\section{Material model}
\label{chap:material}

This section presents several phenomenological material models which are used to describe transversal isotropic materials under consideration of large deformations. In particular, the kinematic foundations as well as the continuum mechanical framework are introduced. The material models are implemented in \aba via the \textit{User Subroutine UMAT}.

\subsection{Kinematics}

\begin{figure}[H]
\centering
\includegraphics[height=3cm]{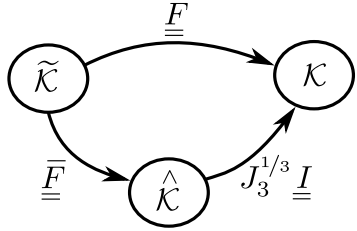}
\caption[]{Commutative diagram.}
\label{fig:schaubild}
\end{figure}

\figref{fig:schaubild} illustrates the underlying kinematics. The deformation gradient $\Ten2F$ maps vectors from the reference configuration $\STAPEL {\mathcal{K}}!^\SLtilde$ to the current configuration $\mathcal{K}$. $\Ten2F$ can be split into its purely volumetric part $\Jhoch{1}{3}\Ten2I$ with $J_3 = \det{\Ten2F}$ and its isochoric part $\Ten2F!^\SLstrich$. This procedure is also referred to as Flory split \citep{Flory.1961}
\begin{equation}
\begin{aligned}
\Ten2F = \Jhoch{1}{3}\Ten2I \pkt \Ten2F!^\SLstrich \qquad\rightarrow\qquad \Ten2F!^\SLstrich = \Jhoch{\minus 1}{3}\Ten2F \qquad.
\end{aligned}
\end{equation}
$\Ten2F!^\SLstrich$ maps vectors from the reference configuration to the intermediate configuration $\STAPEL {\mathcal{K}}!^\wedge$. The right Cauchy-Green tensor of the isochoric deformation is referred to as $\Ten2C!^\SLstrich$
\begin{equation}
\begin{aligned}
\Ten2C!^\SLstrich = \Ten2F!^\SLstrich^T \pkt \Ten2F!^\SLstrich = \Jhoch{\minus 2}{3}\Ten2F^T \pkt \Ten2F = \Jhoch{\minus 2}{3}\Ten2C \qquad.
\end{aligned}
\end{equation}

\subsection{Continuum mechanics framework}

In order to postulate an isotropic energy density $\psi$ (free energy per unit of volume), some invariants are introduced

\begin{equation}
\begin{aligned}
J_1 &= \Ceins \ppkt \Ten2I \qquad ,\\
J_2 &= \frac{1}{2}\left( (\Ceins \ppkt \Ten2I)^2 - \Ceins^2 \ppkt \Ten2I \right) \qquad ,\\
J_3 &= \det{\Ten2F} \qquad .
\end{aligned}
\end{equation}

The invariants $J_i$ with $i=1,2,3$ correspond to the isotropic part of the material model. Note that the Flory split is used, i.e. the invariants are applied to the purely volumetric and the isochoric parts respectively. In order to consider the anisotropic part, the rank 1 structural tensor (cf. \citep{Schroder.2003}) is introduced
\begin{equation}
\begin{aligned}
\A = \Vek{a}!^\SLtilde \dya \Vek{a}!^\SLtilde \qquad .
\end{aligned}
\end{equation}
The vector $\Vek{a}!^\SLtilde$ describes the orientation of the fibres in the reference configuration. In accordance to the concept of the integrity bases (the reader is referred to \citep{Smith.1965, Spencer.1961, Spencer.1958, Spencer.1962, Spencer.1965}), mixed invariants are formulated. Following the argumentation presented in \citep{Sansour.2008} and \citep{Helfenstein.2010}, the mixed invariants $I_4,I_5$ are introduced
\begin{align}
I_4 &= \C \ppkt \A \qquad, \\
I_5 &= \C^2 \ppkt \A \qquad.
\end{align}
In contrast to this, some authors \citep{Holzapfel.2001, Andriyana.2010} define the mixed invariants using the isochoric part of the deformation
\begin{align}
J_4 &= \Ceins \ppkt \A \qquad, \\
J_5 &= \Ceins^2 \ppkt \A \qquad.
\end{align}
Assuming decoupled effects of the isotropic-isochoric, the isotropic-volumetric and the anisotropic parts of the energy density, $\psi$ reads in general
\begin{equation}
\begin{aligned}
\psi &= \psi_{\mathrm{isochoric}}(J_1,J_2) + \psi_{\mathrm{volumetric}}(J_3) + \psi_{\mathrm{anisotropic}}(I_4, I_5) \qquad ,
\end{aligned}
\label{eq:psi_allgemein_I}
\end{equation}
and
\begin{equation}
\begin{aligned}
\psi &= \psi_{\mathrm{isochoric}}(J_1,J_2) + \psi_{\mathrm{volumetric}}(J_3) + \psi_{\mathrm{anisotropic}}(J_4, J_5) \qquad ,
\end{aligned}
\label{eq:psi_allgemein_J}
\end{equation}
respectively. The Clausius-Planck inequality for hyperelastic materials without change of tem\-pera\-ture reduces to the identity
\begin{equation}
\begin{aligned}
\frac{1}{2}\Ten2T!^\SLtilde \ppkt \C!^\SLdreieck - \STAPEL {\psi}!^\SLpunkt = 0 \qquad.
\end{aligned}
\label{eq:CPU}
\end{equation}
The insertion of the energy density \eqref{eq:psi_allgemein_I} or \eqref{eq:psi_allgemein_J} into \eqref{eq:CPU} yields the equation for the stress tensor in the reference configuration
\begin{equation}
\begin{aligned}
\Ttil = 2\frac{\partial \psi}{\partial \C} \qquad.
\end{aligned}
\label{eq:Ttil_allgemein}
\end{equation}
Here, the derivative is understood as a derivative with respect to a symmetric tensor (cf., for example, \citep{Shutov.2008}). The first Piola-Kirchhoff stress tensor $\Ten2T$ as well as the Cauchy stress tensor $\Ten2\sigma$ can be obtained with the help of the following relations
\begin{align}
\Ten2T &= \Ttil \pkt \F^T \qquad, \\
\Ten2\sigma &= \frac{1}{J_3}\F \pkt \Ttil \pkt \F^T \qquad. \label{eq:cauchy}
\end{align}
The Lagrangian stiffness tensor reads
\begin{align}
\Mstiff &= 4\frac{\partial^2 \psi}{\partial\C \partial \C} \qquad. \label{eq:M_allgemein}
\end{align}
According to \citep{Ihlemann.2014}, the commercial FEM software \aba requires the stiffness tensor
\begin{equation}
\frac{1}{J_3}\STAPEL k!_\SLstrich!_\SLstrich!_\SLstrich!_\SLstrich^* = 
\frac{1}{J_3}\left(\left(\F\dya\F^T\right)^{\mathrm{S_{24}}} \ppkt \Mstiff \ppkt \left(\F^T\dya\F\right)^{\mathrm{S_{24}}}\right)
+ 2\left(\Ten2\sigma \dya \Ein\right)^\mathrm{S_{24}}   \qquad,
\end{equation}
which can be computed using the Lagrangian stiffness tensor $\Mstiff$.

\subsection{Concrete ansatz for the free energy density}

\subsubsection{Energy density with mixed invariant $I_4$}

To be definite, the following ansatz for $\psi$ is considered
\begin{equation}
\begin{aligned}
\psi_1 = \frac{G}{2}(J_1-3) + \frac{K}{2}(J_3-1)^2 + \frac{E_F}{6}\left(I_4 + \frac{2}{\sqrt{I_4}} - 3\right) \qquad ,
\end{aligned}
\label{eq:psi1}
\end{equation}
where $G$ is the shear modulus, $K$ is the bulk modulus and $E_F$ is the fibre's Young's modulus. \eqref{eq:psi1} corresponds to an extension of the NeoHooke law and can be found, for example, in \citep{Guo.2007b}. Using \eqref{eq:Ttil_allgemein}, the second Piola-Kirchhoff tensor reads
\begin{equation}
\begin{aligned}
\Ttil = \left\{G\Ceins^D + KJ_3(J_3-1)\Ein\right\}\pkt\Cinv + \frac{1}{3}E_F\left\{1 - I_4^{\minus\nicefrac{3}{2}}\right\}\A \qquad. \label{eq:Ttil1}
\end{aligned}
\end{equation}
According to \eqref{eq:M_allgemein}, the Lagrangian stiffness tensor is defined by
\begin{equation}
\begin{aligned}
\Mstiff &= \frac{2}{3}G\left\{-\Jhoch{-2}{3}(\Ein\dya\Cinv + \Cinv\dya\Ein) + \frac{1}{3}J_1\Cinv\dya\Cinv + J_1\left(\Cinv\dya\Cinv\right)^{\mathrm{S}_{24}}\right\} \\
&\quad+ 2K\left\{\frac{1}{2}J_3(J_3-1)\Cinv\dya\Cinv - J_3(J_3-1)\left(\Cinv\dya\Cinv\right)^{\mathrm{S}_{24}} \right\} \\
&\quad+ E_FI_4^{\nicefrac{-5}{2}}\A\dya\A \qquad. \label{eq:M1}
\end{aligned}
\end{equation}

\subsubsection{Energy density with mixed invariant $J_4$}

Similarly to \eqref{eq:psi1}, $\psi_2$ can be reformulated with the mixed invariant $J_4$ instead of $I_4$
\begin{equation}
\begin{aligned}
\psi_2 = \frac{G}{2}(J_1-3) + \frac{K}{2}(J_3-1)^2 + \frac{E_F}{6}\left(J_4 + \frac{2}{\sqrt{J_4}} - 3\right) \qquad.
\end{aligned}
\label{eq:psi2}
\end{equation}
Thus, the anisotropic part is only defined using the isochoric part of the deformation. The second Piola-Kirchhoff stress tensor then reads
\begin{equation}
\begin{aligned}
\Ttil = \left\{G\Ceins^D + KJ_3(J_3-1)\Ein\right\}\pkt\Cinv + \frac{1}{3}E_F\left\{1 - J_4^{\minus\nicefrac{3}{2}}\right\}\left(\A\pkt\Ceins\right)^D \pkt \Cinv \qquad .
\end{aligned}
\end{equation}
The Lagrangian stiffness tensor follows with
\begin{equation}
\begin{aligned}
\Mstiff &= \frac{2}{3}G\left\{-\Jhoch{-2}{3}(\Ein\dya\Cinv + \Cinv\dya\Ein) + \frac{1}{3}J_1\Cinv\dya\Cinv + J_1\left(\Cinv\dya\Cinv\right)^{\mathrm{S}_{24}}\right\} \\
&\quad+ 2K\left\{\frac{1}{2}J_3(J_3-1)\Cinv\dya\Cinv - J_3(J_3-1)\left(\Cinv\dya\Cinv\right)^{\mathrm{S}_{24}} \right\} \\
&\quad+ E_FJ_4^{\nicefrac{-5}{2}}\left\{\frac{1}{9}J_4^2\Cinv\dya\Cinv - \frac{1}{3}(\Cinv\dya\A + \A\dya\Cinv) + \Jhoch{-4}{3} \A\dya\A \right\} \\
&\quad+ \frac{2}{3}E_F(1-J_4^{\nicefrac{-3}{2}})\left\{ \frac{1}{9}J_4\Cinv\dya\Cinv - \frac{1}{3}\Jhoch{-2}{3}(\Cinv\dya\A + \A\dya\Cinv) \right.\\
&\qquad+ \left.\frac{1}{3}J_4\left(\Cinv\dya\Cinv\right)^{\mathrm{S}_{24}} \right\} \qquad.
\end{aligned}
\end{equation}

\subsubsection{Convex combination according to the fibre volume fraction}

Another ansatz for the energy density stems from the convex combination of the single parts according to the fibre volume fraction. Such an assumptions can be found, for example, in \citep{Sansour.2008} and is usually applied to materials with endless fibres. However, some authors (\citep{Andriyana.2010}) make use of it for short fibre reinforced materials too. The corresponding relation reads
\begin{align}
\psi_3 &= (1-f_{\mathrm{vol}})\psi_{\mathrm{matrix}} + f_{\mathrm{vol}}\psi_{\mathrm{fibre}} \qquad, \\
\begin{split}
\psi_3 &= (1-f_{\mathrm{vol}})\left\{\frac{G_M}{2}(J_1-3) + \frac{K_M}{2}(J_3-1)^2\right\} \\
       &\quad+ f_{\mathrm{vol}}\left\{\frac{G_F}{2}(J_1-3) + \frac{K_F}{2}(J_3-1)^2 + \frac{E_F}{6}\left(I_4 + \frac{2}{\sqrt{I_4}} - 3\right)\right\} \qquad, \label{eq:psi3}
\end{split}					
\end{align}
where $0\le f_\mathrm{vol} \le 1$ stands for the fibre volume fraction.
The equations for the stress and the stiffness correspond to \eqref{eq:Ttil1} and \eqref{eq:M1} and only differ in the scalar factors $(1-f_{\mathrm{vol}})$ in front of the material parameters $G_M,K_M$ and $f_{\mathrm{vol}}$ in front of $G_F,K_F,E_F$.

\section{Modelling}
\label{chap:modelling}

In this section, a representative volume element (RVE) of a short fibre reinforced material is introduced. The RVE contains several unidirectional fibres which are randomly distributed in the matrix material. The RVE's shape is restrained to a very simple geometry, i.e. a cuboid. This cuboid is discretised by identical finite elements, which allow a very fast and simple meshing of the RVE during the FEM's preprocessing. This rough approximation of the fibre's geometry is sufficient to cover the global characteristics of the RVE, however, local effects like failure cannot be considered in a proper way. Several parameters can be adjusted in order to represent a broad variety of materials (see \tabref{tab:rve_parameters}). 

\begin{table}[H]
\centering
\caption[]{Parameters for the modelling of the RVE.}
\label{tab:rve_parameters}
\begin{tabular}{cc}
\toprule
Description & symbol \\
\midrule
fibre volume fraction & $f_{\mathrm{vol}}$ \\
minimum distance between fibres & $d_{\mathrm{fib}}$ \\
fibres per spatial direction & \textit{FPD} \\
length-to-width-ratio & $\chi$ \\
\bottomrule 
\end{tabular}
\end{table}

\begin{figure}[H]
\centering
\subfloat[coarse mesh]{\includegraphics[width=.48\textwidth]{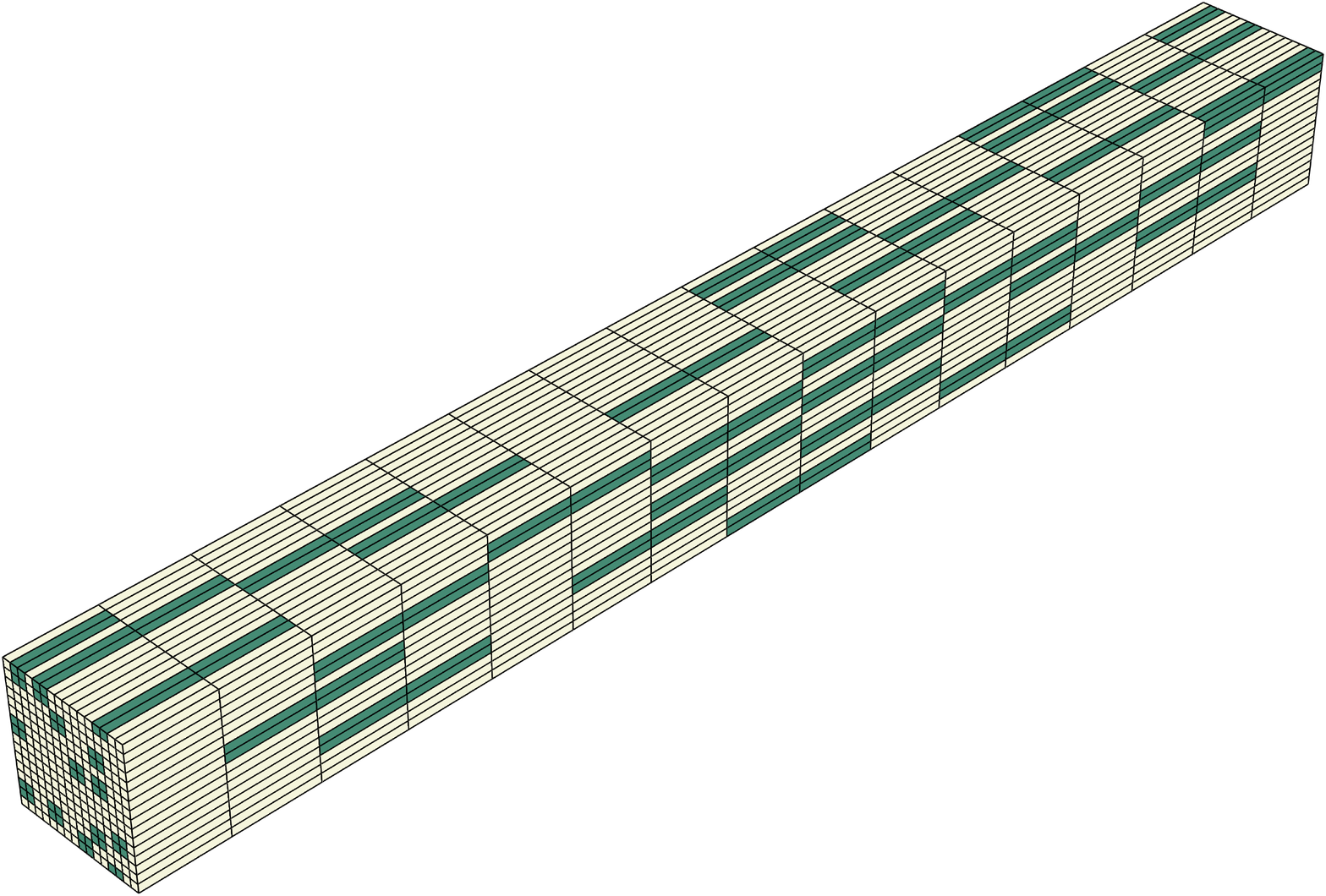}} \quad
\subfloat[fine mesh]{\includegraphics[width=.48\textwidth]{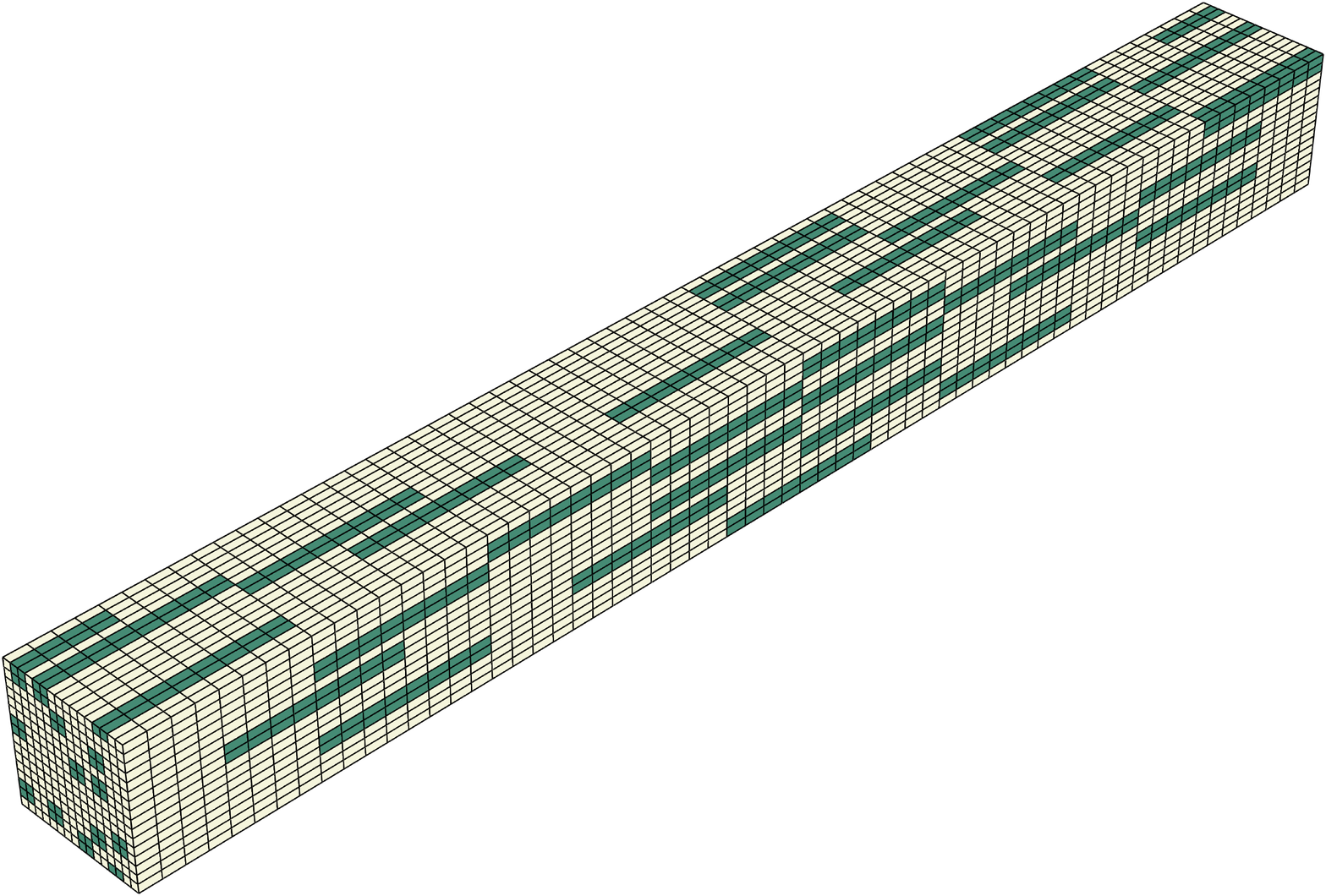}}
\caption[]{Computation models with different meshes built from the same base model.}
\label{fig:calc_models}
\end{figure}

According to these parameters, a base model is generated. The base model remains unchanged in the following. Further modifications with respect to, for example, the degree of the discretisation or the boundary conditions, are done to copies of the base model, the so called computation models. Thus, different simulations based on the same fibre distribution are available and offer reliable comparison. \figref{fig:calc_models} shows different computation models. Note, that their shape resembles a long cuboid more than a regular cube. This originates from the consideration of the parameter \textit{FPD} (average amount of fibres per spatial direction) in combination with the high length-to-width-ratio of the fibres. The parameter \textit{FPD} is required for a stochastically equivalent variation of the RVE's size in section \ref{chap:simulations}. 

\begin{figure}[H]
\centering
\includegraphics[height=3cm]{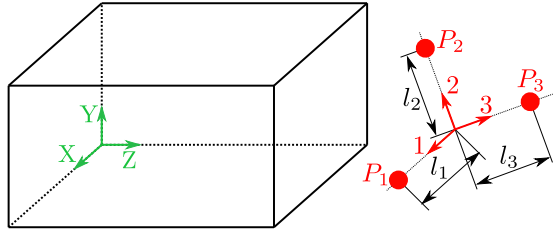}
\caption[]{Illustration of the additional pilot nodes. The elements and nodes of the RVE are defined in the global coordinate system (base vectors $\left.\Vek{e}_a\right|_{a=X,Y,Z}$). The pilot nodes lie in an external, local coordinate system (base vectors $\left.\Vek{e}_i\right|_{i=1,2,3}$).}
\label{fig:pilotknoten}
\end{figure}

Each computation model contains an additional local coordinate system with three auxiliary nodes, the pilot nodes $P_1, P_2, P_3$ (see \figref{fig:pilotknoten}). Those nodes are used to determine the average displacement gradient and first Piola-Kirchhoff stress tensor as explained in section \ref{chap:boundary}. The pilot nodes' distances to the local coordinate system's origin are described by $l_1,l_2,l_3$ and can be chosen arbitrarily. In this work they are set to $l_1 = l_2 = l_3 = 1mm$.

\section{Boundary conditions}
\label{chap:boundary}

Before the RVE is subjected to external loads or displacements, conditions on its boundary have to be defined. The purpose of the RVE is the determination of effective material properties. Therefore, the application of homogeneous boundary conditions appears suitable. There exist several concepts for homogeneous boundary conditions in the literature (see \citep{Lee.1993} and references within). All of them fulfill the Hill condition \citep{Hill.1963} per se and independently from the RVE's size.  

\subsection{Periodic boundary conditions}

\begin{figure}[H]
\centering
\includegraphics[height=5cm]{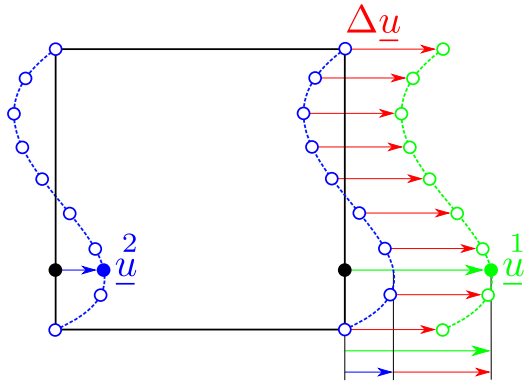}
\caption[]{Principle of periodic boundary conditions. The displacements of the right edge (green) are coupled to the displacements of the left edge (blue). As an average and therefore constant displacement gradient is imposed on the entire surface, the displacement difference $\Delta\Vek{u}$ between right and left edge is constant for every pair of points.}
\label{fig:pbc_skizze}
\end{figure}

This work focuses solely on periodic boundary conditions (PBC). PBC offer a compromise between homogeneous traction and homogeneous strain boundary conditions. They require pairs of points, with position vectors $\Vek{x}!^\SLtilde!^1$ and $\Vek{x}!^\SLtilde!^2$. The PBC couple the displacements of those pairs of points
\begin{equation}
\begin{aligned}
&\Vek{u}!^1 - \Vek{u}!^2 = \Hav \pkt (\Vek{x}!^\SLtilde!^1 - \Vek{x}!^\SLtilde!^2) \qquad, \\
\end{aligned}
\label{eq:PBC}
\end{equation}
with $\Vek{u}$ being the displacement vector and $\Hav = \Fav - \Ten2I$ the average displacement gradient (cf. \citep{Gluge.2012, Gluge.2013}). Thus, the PBC are also referred to as coupled constraints. \figref{fig:pbc_skizze} illustrates the principle of PBC.

In consideration of the periodicity, fibres, which pierce the boundary surface, continue at the opposite surface (see \figref{fig:RVE_final}).

\subsection{Implementation in \aba}

As already mentioned, PBC are coupled constraints between pairs of points. \aba offers the implementation of such constraints in the form of \textit{EQUATIONS} (see \citep{Abaqus}, section 33.2.1), which linearly relate the displacements of nodes. In order to demonstrate the applicability of those to the PBC, \eqref{eq:PBC} is reformed. Firstly, some abbreviations are introduced
\begin{align}
\Vek{u}!^1 - \Vek{u}!^2 &= \Delta\Vek{u} \qquad ; \qquad \Vek{x}!^\SLtilde!^1 - \Vek{x}!^\SLtilde!^2 = \Delta\Vek{x}!^\SLtilde \qquad , \\
\Delta\Vek{u} &= \Hav \pkt \Delta\Vek{x}!^\SLtilde \qquad . \label{eq:PBC_koeff_free}
\end{align}
\aba demands explicit equations for the displacements coefficients, which is why \eqref{eq:PBC_koeff_free} is written in notation with coordinates
\begin{equation}
\Delta u_c \Vek{e}_c = \STAPEL H!^*_{ij}\Vek{e}_i \dya \Vek{e}_j \pkt \Delta\STAPEL x!^\SLtilde_a \Vek{e}_a \qquad. \\
\label{eq:PBC_koeff_allg}
\end{equation}
Note that the coefficients of $\Delta\Vek{u}$ and $\Delta\Vek{x}!^\SLtilde$ refer to the global coordinate system whereas the coefficients of $\Hav$ refer to the local coordinate system. In order to obtain the equation for the single coefficients of $\Delta\Vek{u}$, the base vectors are contracted with each other
\begin{align}
\Delta u_c \Vek{e}_c &= \STAPEL H!^*_{ij} \Delta\STAPEL x!^\SLtilde_a (\Vek{e}_j \pkt \Vek{e}_a)\Vek{e}_i \qquad | \pkt \Vek{e}_b \qquad , \\
\Delta u_b &= \STAPEL H!^*_{ij} \Delta\STAPEL x!^\SLtilde_a (\Vek{e}_j \pkt \Vek{e}_a)(\Vek{e}_i \pkt \Vek{e}_b) \qquad . \label{eq:EQ_cof}
\end{align}
On the left hand side of \eqref{eq:EQ_cof} there are the degrees of freedom (DOF) of the two nodes, which are to be coupled. On the right side there is the distance between those two points in the reference configuration $\Delta\STAPEL x!^\SLtilde_a$ which is a constant value. The scalars $(\Vek{e}_j \pkt \Vek{e}_a)$ and $(\Vek{e}_i \pkt \Vek{e}_b)$ are constant too and relate the directions of both the global and the local coordinate system (see \figref{fig:pilotknoten}). The only unknown values are the coefficients of the average displacement gradient $\STAPEL H!^\SLstrich_{ij}$. Those are determined by
\begin{equation}
\begin{aligned}
\STAPEL H!^*_{ij} = \dfrac{\STAPEL u!^{P_j}_i}{l_j} \qquad .
\end{aligned}
\label{eq:H_koef}
\end{equation}
Thus, all necessary values are known and a system of linear equations can be formulated. \aba generally requires equations in the form
\begin{equation}
\begin{aligned}
a_1\Skal{u}!^1_1 + a_2\Skal{u}!^2_1 + \dots + a_N\Skal{u}!^N_1 &= 0 \qquad ,\\
a_1\Skal{u}!^1_2 + a_2\Skal{u}!^2_2 + \dots + a_N\Skal{u}!^N_2 &= 0 \qquad ,\\ 
a_1\Skal{u}!^1_3 + a_2\Skal{u}!^2_3 + \dots + a_N\Skal{u}!^N_3 &= 0 \qquad ,\\
\end{aligned}
\label{eq:EQ_allgemein}
\end{equation}
where $a_i$ are scalar factors and $N$ is the number of involved DOF. The average displacement gradient is defined by three pilot nodes with three DOF each. Together with the two DOF from the two nodes to be coupled, \eqref{eq:EQ_cof} yields a total number of DOF, $N=11$. Hence, the equation for each $b \in [1,3]$ reads
\begin{equation}
\begin{aligned}
&a_{1}\Skal{u}!^{1}_b + a_{2}\Skal{u}!^{2}_b&
  +& a_{11}\Skal{u}!^{P_1}_1 &+& a_{12}\Skal{u}!^{P_2}_1 &+& a_{13}\Skal{u}!^{P_3}_1& \\
&&+& a_{21}\Skal{u}!^{P_1}_2 &+& a_{22}\Skal{u}!^{P_2}_2 &+& a_{23}\Skal{u}!^{P_3}_2& \\
&&+& a_{31}\Skal{u}!^{P_1}_3 &+& a_{32}\Skal{u}!^{P_2}_3 &+& a_{33}\Skal{u}!^{P_3}_3& = 0 \qquad,\\
\end{aligned}
\label{eq:EQ_konkret}
\end{equation}
with
$$
a_1 = 1, \quad a_2 = -1, \quad a_{ij} = -\frac{1}{l_j}\sum_{a=x}^z (\Skal{x}!^\SLtilde!^1_a - \Skal{x}!^\SLtilde!^2_a)(\Vek{e}_j \pkt \Vek{e}_a)(\Vek{e}_i \pkt \Vek{e}_b) \qquad .
$$

\subsection{Consideration of tilted fibres}

\begin{figure}[H]
\centering
\includegraphics[height = 4cm]{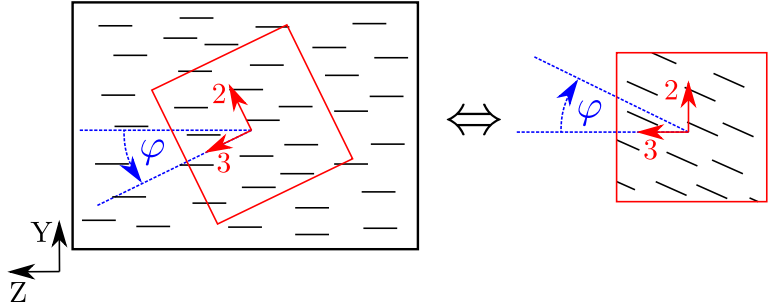}
\caption[]{The local coordinate system is rotated about the global X-axis by the fibre angle $\varphi$ (left). Therefore, the imposed deformations accord to tilted fibres with the fibre angle $-\varphi$ (right). }
\label{fig:faserwinkel}
\end{figure}

\eqref{eq:EQ_cof} holds for arbitrary base vectors $\Vek{e}_i$ and $\Vek{e}_j$. Therefore the local coordinate system does not need to be aligned parallel to the global coordinate system. Nonetheless, the rotation of the local coordinates system bears some consequences as it changes the physical meaning of the deformation. Whereas the direction of the fibres stays constant (parallel to base vector $\Vek{e}_Z$), the deformation points in the direction corresponding to the local coordinate system. Thus, an easy way to consider tilted fibres is provided (see \figref{fig:faserwinkel}). The local base vector $\Vek{e}_1$ is set to lie parallel to the global base vector $\Vek{e}_X$, i.e. $(\Vek{e}_1 \pkt \Vek{e}_X) = 1$. That is why from now on a rotation about the X-axis is referred to as fibre angle $\varphi = \sphericalangle(\Vek{e}_Z;\Vek{e}_3)$.

\subsection{Average strain and stress}

The examination of the average strain and stress is done with the help of the average deformation gradient $\Fav$ and the average first Piola-Kirchhoff stress tensor $\Tav$. Both quantities are chosen due to their easy determination and because they are power conjugate. With
\begin{equation}
\begin{aligned}
\Fav = \Hav + \Ten2I \qquad ,
\end{aligned}
\label{eq:Fav}
\end{equation}
and \eqref{eq:H_koef} the deformation gradient can be directly related to the pilot nodes' displacements. Furthermore the time derivative of $\Fav$ is given by
\begin{equation}
\Fav!^\SLdreieck = \Hav!^\SLdreieck + \Ten2I!^\SLdreieck = \Hav!^\SLdreieck \qquad,
\end{equation}
\begin{equation}
\begin{aligned}
\STAPEL H!^*!^\SLdreieck_{ij} = \dfrac{\STAPEL v!^{P_j}_i}{l_j} \qquad, 
\end{aligned}
\label{eq:Hzeit}
\end{equation}
with $\STAPEL v!^{P_j}_i$ being the coefficient $i$ of the velocity at pilot node $P_j$. The stress power $P_s$ is defined as integral over the volume $\STAPEL V!^\SLtilde$ in the reference configuration
\begin{equation}
P_s = \int \Ten2T \ppkt \Ten2F!^\SLdreieck d\STAPEL V!^\SLtilde = \int \Ten2T \ppkt \Ten2H!^\SLdreieck d\STAPEL V!^\SLtilde = \STAPEL V!^\SLtilde\Tav \ppkt \Hav!^\SLdreieck
    = \STAPEL V!^\SLtilde \STAPEL T!^*_{ab} \STAPEL H!^*!^\SLdreieck_{ba}  \qquad.
\label{eq:stresspower}		
\end{equation}
\eqref{eq:stresspower} provides the definition of a stress tensor $\Tav$ according to \eqref{eq:T_koef}. Note that the stress power becomes a sum of nodal forces multiplied with corresponding nodal velocities (see \eqref{eq:T_proof})
\begin{equation}
\Skal{T}!^*_{ij} = \frac{\Skal{Q}!^{P_i}_j \pkt l_i}{\Skal{V}!^\SLtilde} \qquad .
\label{eq:T_koef}
\end{equation}
Here $\Skal{Q}!^{P_i}_j$ are the coefficients of the nodal forces at the pilot node $P_i$. Inserting \eqref{eq:Hzeit} and \eqref{eq:T_koef} into \eqref{eq:stresspower} results into
\begin{equation}
P_s = \STAPEL V!^\SLtilde \pkt \frac{\Skal{Q}!^{P_i}_j \pkt l_i}{\Skal{V}!^\SLtilde} \pkt \dfrac{\STAPEL v!^{P_j}_i}{l_j} = \Skal{Q}!^{P_i}_j \pkt \STAPEL v!^{P_j}_i \qquad. \label{eq:T_proof}
\end{equation}
With \eqref{eq:H_koef}, \eqref{eq:Fav} and \eqref{eq:T_koef} a full set of average boundary conditions can be prescribed, be it for deformation or stress. In order to achieve a static equilibrium, nine coefficients $\Skal{T}!^*_{ab}$ or $\Skal{F}!^*_{ba}$ must be defined. Furthermore, three of the following deformations must be prescribed in order to avoid rigid body rotations
\begin{equation}
\begin{aligned}
	&\FavKoef{12} &\text{or} &&\FavKoef{21} \qquad, \\
	&\FavKoef{13} &\text{or} &&\FavKoef{31} \qquad, \\
	&\FavKoef{23} &\text{or} &&\FavKoef{32} \qquad. \\
\end{aligned}
\end{equation}


\section{Generation of synthetic data}
\label{chap:simulations}

The RVE introduced in section \ref{chap:modelling} is now used to generate synthetic experimental data. Therefore, the PBC together with different combinations of stress and deformation boundary conditions ($\Skal{T}!^*_{ab}$ and $\Skal{F}!^*_{ab}$) are applied. First, some investigations concerning the average stress as well as the convergence are presented.

\subsection{Verification of the average stress}
\begin{figure}[H]
\centering
\import{simulations/}{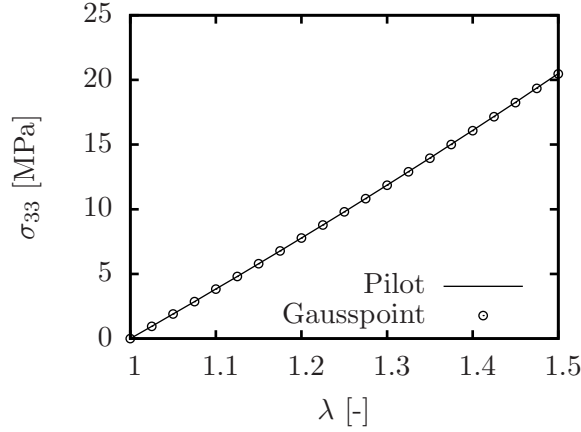}
\caption[]{Verification of average stress obtained by evaluating the pilot's nodal forces. The results are compared to the stress values, which were received by averaging all Gauss points.}
\label{fig:spannungsvergleich}
\end{figure}

In order to verify \eqref{eq:T_koef}, a RVE made up of 25\% fibres is subjected to uniaxial tension with a stretch of $\lambda = 1.5$ in the direction of fibres, i.e. $\varphi=0^\circ$. Then the resulting homogenous Cauchy stress is calculated using both \eqref{eq:T_koef} together with \eqref{eq:cauchy} and the average over all Gauss point (GP) values. \figref{fig:spannungsvergleich} shows the comparison of both methods. Both curves are almost identical and therefore indicate the correctness of \eqref{eq:T_koef}.

\subsection{Convergence}

Here, the influence of the RVE's size and mesh density is examined. Toward this end, the norm of the Eulerian stiffness is chosen as a measure for comparison. The stiffness $\Cstiff$ relates the Cauchy stress $\Ten2\sigma$ and the linearised strain $\Ten2\varepsilon!^{\mathrm{lin}}$
\begin{equation}
\begin{aligned}
\Ten2\sigma = \Cstiff \ppkt \Ten2\varepsilon!^{\mathrm{lin}} \qquad .
\end{aligned}
\end{equation}
The coefficients follow with
\begin{equation}
\begin{aligned}
C_{abcd} = \frac{\partial\sigma_{ab}}{\partial\varepsilon_{cd}} \qquad .
\end{aligned}
\end{equation}

\subsubsection{Size of the RVE}

The parameter \FPD \ determines the average number of fibres per spatial direction and hence allows a stochastically equivalent variation of the RVE's size. It is increased from 1 to 4 whereas all other parameters are kept constant. For each value of \FPD \ ten computation models are created. Thus, there result ten stiffness tensors for ten different fibre distributions. Those ten stiffness tensors are averaged. In order to compare the averaged stiffness tensors of each value of \FPD, the relative norm of the difference tensor is computed. The most accurate result is assumed for the highest number of fibres and therefore biggest RVE. Then the relative norm $\|\Delta\Cstiff!^i_{\mathrm{rel}}\|$ reads
\begin{equation}
\begin{aligned}
\|\Delta\Cstiff!^i_{\mathrm{rel}}\| = \frac{\|\Cstiff!^i - \Cstiff!^4\|}{\|\Cstiff!^4\|}\pkt100\% \qquad , \qquad i=1,2,3,4 \qquad .
\end{aligned}
\end{equation}
As the mean value of the stiffness tensors does not deliver any information about the fluctuations, the coefficient of variation is additionally computed and can be found in \tabref{tab:COV}. \figref{fig:norm_fpd} shows the course of the relative norm over the parameter \FPD. The relative norm of the difference between the stiffness tensors is already very small for $FPD=1$. The corresponding coefficient of variation is also already very small with only 1.34$\%$. Due to the fact that an increase of \FPD \ results in a significant increase of the computation time, $FPD=1$ is chosen in the following simulations.
\begin{minipage}[t]{.50\textwidth}
\begin{figure}[H]
\centering
\import{simulations/}{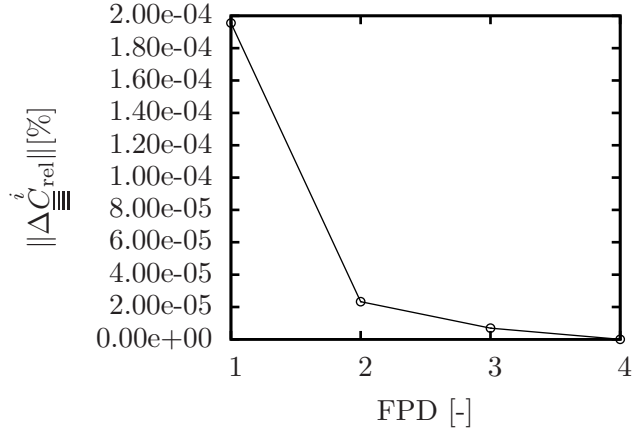}
\caption[]{Relative norm of stiffness over parameter \FPD.}
\label{fig:norm_fpd}
\end{figure}
\end{minipage}\hfill
\begin{minipage}[t]{.45\textwidth}
\begin{table}[H]
\centering
\setlength\belowcaptionskip{2pt}
\caption[]{Coefficient of variation of stiffness tensor's norm for different values of \textit{FPD}.}
\label{tab:COV}
\begin{tabular}{cc}
\toprule
\textit{FPD} & COV [$\%$] \\
\midrule
1 & 1.3415 \\
2 & 0.2577 \\
3 & 0.1840 \\
4 & 0.1148 \\
\bottomrule 
\end{tabular}
\end{table}
\end{minipage}

\subsubsection{Mesh density}

\begin{figure}[H]
\centering
\includegraphics[width=.6\textwidth]{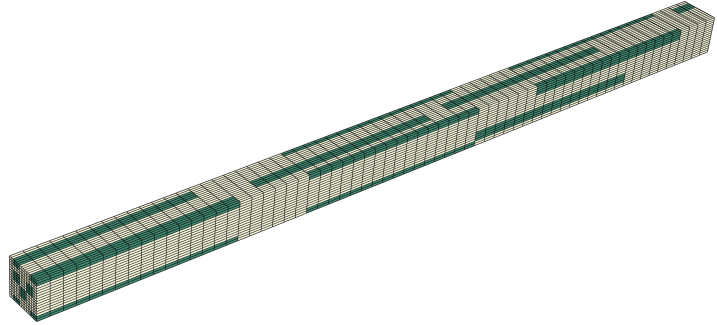}
\caption[]{RVE for the simulations. The modelling parameters are chosen so that the material behaviour converges sufficiently. Fibres, which pierce the surface, continue at the opposite side.}
\label{fig:RVE_final}
\end{figure}

A similar comparison is done for the mesh density, meaning again the relative norm of the difference of the stiffness tensors will be examined for different mesh densities. This time only one base model but different computation models with different meshes come into operation. The mesh density is influenced by the number of elements per fibre in transversal ($n_\bot$) and longitudinal ($n_\parallel$) direction. Additionally, the computation time is recorded because an increase of the element number significantly affects the simulation time. Similarly to the relative norm of the stiffness difference, a relative measure of time $\Skal{t}!^i_{\mathrm{rel}}$ is calculated. Here the model with the fewest elements is used as a reference 
\begin{equation}
\begin{aligned}
\Skal{t}!^i_{\mathrm{rel}} = \frac{\Skal{t}!^i}{\Skal{t}!^{\mathrm{min}}} \qquad .
\end{aligned}
\end{equation}
In contrast to the analysis of the RVE's size, the convergence measures depend on two instead of one parameters. This is why both measures are plotted over several pairs $(n_\bot , n_\parallel)$.
\begin{figure}[htb]
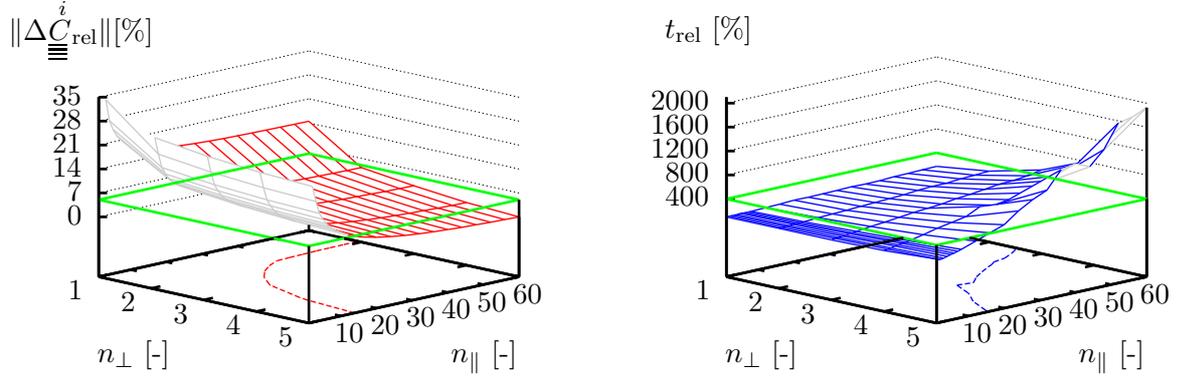

\centering
\subfloat{ \import{simulations/}{pic_Norm_MD} }
\subfloat{ \import{simulations/}{pic_Zeit_MD} }
\caption[]{Relative norm of the stiffness difference (left) and relative computation time (right) for different combinations of element numbers per fibre in transversal and longitudinal direction. The green planes mark the acceptable values. The dashed lines belong to the mesh densities which correspond to these values.}
\label{fig:norm_mesh}
\end{figure}
\figref{fig:norm_mesh} illustrates the influence of the mesh density on the convergence and the computation time. The graphs reveal the significant influence of the mesh density on the convergence. The coarsest mesh (not plotted for the purpose of clarity) results in a norm of the stiffness approximately 70\% higher than the converged one. The difference of the stiffness decreases with finer meshes. Apparently, the lack of a sufficient number of elements causes a stiffening of the simulated material. The flat slope of the surface plot at high values of $n_\bot$ and $n_\parallel$ indicates almost converged material behaviour. It can be seen that the number of elements in transversal direction plays a minor role. The same can be observed for the ratio
$$\frac{n_\parallel}{n_\bot} \qquad. $$
More important is the absolute number of elements as well as the number of elements in longitudinal direction. The plot of the computation time displays an opposite trend. With increasing numbers of elements, the needed time for the computation increases too. 
A relative difference of the stiffness tensors of $5\%$ is considered as converged. All combinations of $n_\bot$ and $n_\parallel$ to the right of the dashed line in the left graph fulfill this convergence criterion. However, more accurate results require higher computation times. The combinations of $n_\bot$ and $n_\parallel$ to the left of the dashed line in the right graph need less than 400\% of the computation time of the coarsest mesh. A good compromise between accuracy and low computation time can be found at $n_\bot=4$ and $n_\parallel=16$. Such a RVE can be seen in \figref{fig:RVE_final}.

\subsection{Synthetic experimental data}
\label{chap:examples}

In the following, four different simulations with the RVE from \figref{fig:RVE_final} are performed. Thus, the RVE's capability to consider different fibre angles is demonstrated. The corresponding experimental parameters can be found in \tabref{tab:setup_simulation}. The boundary conditions are listed in \tabref{tab:BC_simulation}.

\begin{table}[H]
\centering
\caption[]{Parameter setup of the simulations.}
\label{tab:setup_simulation}
\begin{tabular}{lll}
\toprule
\textbf{Parameter} & \multicolumn{2}{l}{\textbf{Value(s)}} \\
\midrule
$f_{\mathrm{vol}}$ & \multicolumn{2}{l}{$25\%$} \\
$\chi$ & \multicolumn{2}{l}{$20$} \\
\FPD & \multicolumn{2}{l}{$1$} \\
$\varphi$ & \multicolumn{2}{l}{$\{0^\circ, 15^\circ, 30^\circ, 45^\circ, 60^\circ, 75^\circ, 90^\circ\}$} \\
\midrule
\textbf{Material} & \textbf{Type} & \textbf{Parameter} \\
\midrule
fibre & NeoHooke\footnotemark[1] & $\mathrm{C}_{10} = 25\quad,\quad\mathrm{D}_{1} = 0.018467 $ \\
matrix & NeoHooke\footnotemark[1] & $\mathrm{C}_{10} = 1\quad,\quad\mathrm{D}_{1} = 0.2 $ \\
\bottomrule 
\end{tabular}
\end{table}
\footnotetext[1]{Those are parameters required by \aba. They correspond to $C_{10}=\frac{G}{2}$ and $D_{1}=\frac{2}{K}$}

\begin{table}[htb]
\centering
\caption[]{Boundary conditions of the four simulations.}
\label{tab:BC_simulation}
\begin{tabular}{lll}
\toprule
Simulation & Boundary Condition & Limits \\
\midrule
\simu{1}, tension/compression & 
$
\begin{bmatrix}
\TavKoef{11} = 0 & \FavKoef{12} = 0 & \FavKoef{13} = 0 \\
\FavKoef{21} = 0 & \TavKoef{22} = 0 & \FavKoef{23} = 0 \\
\FavKoef{31} = 0 & \FavKoef{32} = 0 & \FavKoef{33} = \lambda \\
\end{bmatrix}
$ &
$\lambda \in [0.818,1.2]$ \\[5ex]
\simu{2}, simple shear &
$
\begin{bmatrix}
\FavKoef{11} = 1 & \FavKoef{12} = 0 & \FavKoef{13} = 0 \\
\FavKoef{21} = 0 & \FavKoef{22} = 1 & \FavKoef{23} = \kappa \\
\FavKoef{31} = 0 & \FavKoef{32} = 0 & \FavKoef{33} = 1 \\
\end{bmatrix}
$ &
$\kappa \in [-0.2,0.2]$ \\[5ex]
\simu{3}, purely volumetric deformation &
$
\begin{bmatrix}
\FavKoef{11} = \lambda & \FavKoef{12} = 0 & \FavKoef{13} = 0 \\
\FavKoef{21} = 0 & \FavKoef{22} = \lambda & \FavKoef{23} = 0 \\
\FavKoef{31} = 0 & \FavKoef{32} = 0 & \FavKoef{33} = \lambda \\
\end{bmatrix}
$ &
$\lambda \in [0.818,1.2]$ \\[5ex]
\simu{4}, confined compression &
$
\begin{bmatrix}
\FavKoef{11} = 1 & \FavKoef{12} = 0 & \FavKoef{13} = 0 \\
\FavKoef{21} = 0 & \FavKoef{22} = 1 & \FavKoef{23} = 0 \\
\FavKoef{31} = 0 & \FavKoef{32} = 0 & \FavKoef{33} = \lambda \\
\end{bmatrix}
$ &
$\lambda \in [0.818,1.0]$ \\
\bottomrule 
\end{tabular}
\end{table}

\begin{figure}[H]
\centering
\subfloat[Cauchy stress\label{sub:zugspannung}]{\footnotesize{\import{simulations/}{pic_Zug}}}
\subfloat[sideways motion\label{sub:knicken}]{\includegraphics[width=.49\textwidth]{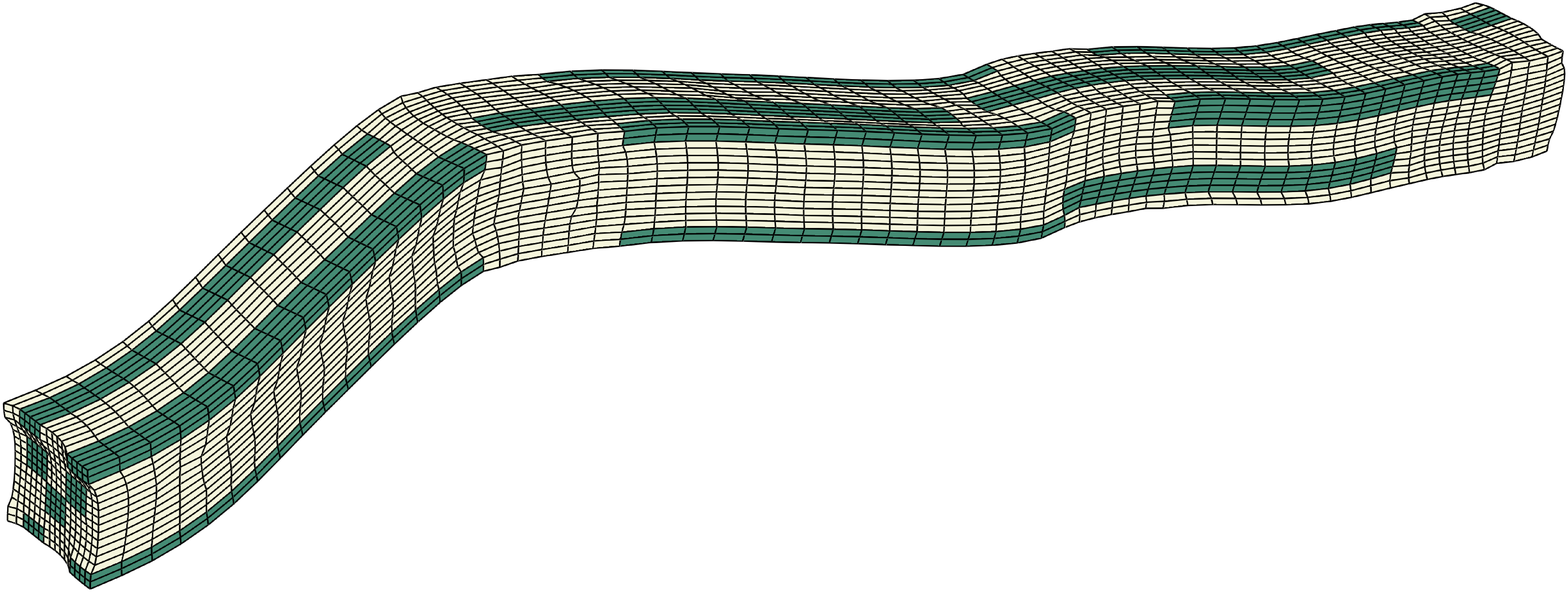}}
\caption[]{Cauchy stress over strain during tension/compression \subref{sub:zugspannung} and buckling of the RVE at values $\lambda \le 0.9$ \subref{sub:knicken}.}
\label{fig:RVE_zug}
\end{figure}

\figref{fig:RVE_zug} shows the course of $\sigma_{33}$ over the stretch $\lambda$ in a tension/compression test. It can be seen that an increase of fibre angle $\varphi$ results into a decrease of the stiffness. The curves of fibre angles $\varphi \le 30^\circ$ exhibit a kink at small stretches $\lambda \le 0.9$. This behaviour stems from the RVE's ability to answer with a sideways motion, due to an unsymmetric fiber distribution. Because of the periodic boundary conditions, the homogenised material behaviour inherits the same tendency to sideways motion. However, in a real material there are patterns with different fiber distributions and therefore tendencies to different sideways motions, which compensate each other. That is why the RVE's sideways motion is an artificial effect.

\begin{figure}[H]
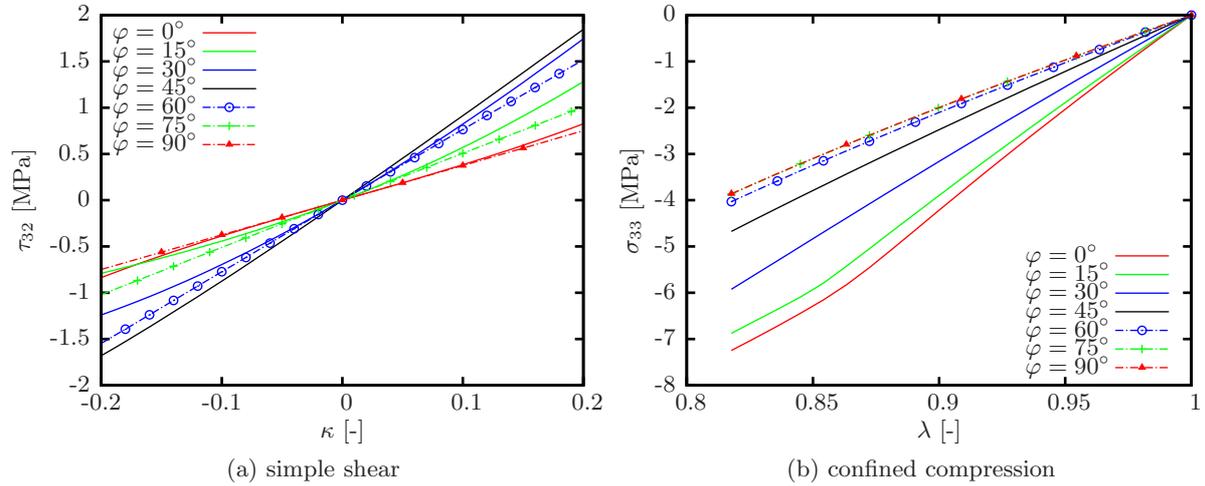

\centering
\subfloat[simple shear\label{sub:shear}]{\footnotesize{\import{simulations/}{pic_Scherung}}}
\subfloat[confined compression\label{sub:compression}]{\footnotesize{\import{simulations/}{pic_DruckDilatation}}}
\caption[]{Cauchy stress over strain during simple shear \subref{sub:shear} and confined compression \subref{sub:compression}.}
\label{fig:RVE_scherung_druckdilatation}
\end{figure}

The evaluation of simple shear (see \figref{sub:shear}) reveals a clear anisotropic impact by the fibres. The stress curves grow linearly and in some cases non linearly. This fact correlates with the different deformations the fibres exhibit at the single fibre angles and values of $\kappa$. The material response for $\varphi = 0^\circ, 90^\circ$ is the same at both directions of shear. In contrast to this, with $\varphi = 15^\circ, 30^\circ, 45^\circ$, the fibres experience  lengthening at positive values of $\kappa$ and shortening at negative values of $\kappa$. Thus, a shear anisotropy is visible. However, that is not the case for $\varphi = 60^\circ, 75^\circ$. The confined compression results (see \figref{sub:compression}) resemble tension/compression with $\lambda \le 1.0$ but exhibit a higher stiffness since the RVE's volume is additionally changed.

\begin{figure}[H]
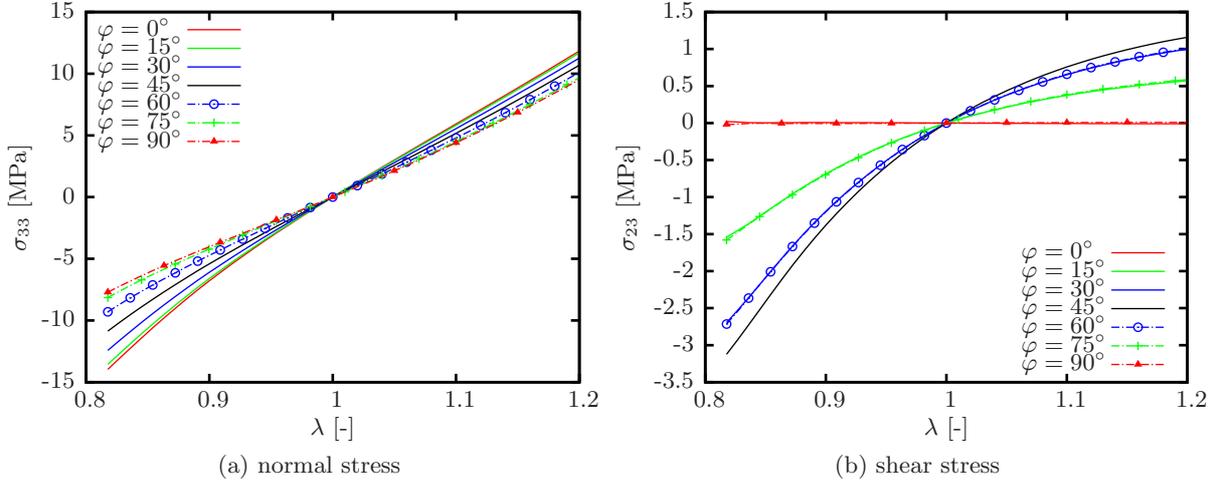

\centering
\subfloat[normal stress\label{sub:dil_normal}]{\footnotesize{\import{simulations/}{pic_Dilatation_S33}}}
\subfloat[shear stress\label{sub:dil_shear}]{\footnotesize{\import{simulations/}{pic_Dilatation_S23}}}
\caption[]{Cauchy stress over strain during purely volumetric deformation.}
\label{fig:RVE_dilatation}
\end{figure}

Lastly, the stresses of purely volumetric deformation at different fibre angles are compared in \figref{fig:RVE_dilatation}. Similarly to tension/compression, the stress $\sigma_{33}$ decreases with growing fibre angles (see \figref{sub:dil_normal}). In accordance to the boundary conditions, which prohibit shear deformation, there are shear stresses (see \figref{sub:dil_shear}), meaning a purely volumetric deformation induces a non-hydrostatic stress tensor. The shear stresses align symmetrically around $\varphi = 45^\circ$ and vanish at $\varphi = 0^\circ, 90^\circ$.

\section{Parameter identification}
\label{chap:identification}

In this section, the material parameters $G,K,E_F$ from the three energy densities \eqref{eq:psi1},\eqref{eq:psi2} and \eqref{eq:psi3} are fitted to the synthetic data obtained in the previous section. First, the methodology of the identification will be shortly explained. Then the predictions of the hyperelastic models with the identified parameters will be compared to the data obtained by the RVE.

\subsection{The optimisation procedure}

The optimisation is performed with the help of \matlab using the built-in function \textit{lsqnonlin}. During each iteration, \matlab invokes \aba with a new set of material parameters. \aba then computes the values of the target function at a single Gauss point using the corresponding boundary conditions. Subsequently, the residuum vector is defined by the difference of the target values and the synthetic data. The starting parameters as well as the parameters of the RVE materials can be found in \tabref{tab:start}.
\begin{table}[H]
\centering
\caption[]{Initial parameters (left), shear and bulk modulus of fibres and matrix in the RVE (right).}
\label{tab:start}
\begin{tabular}{cccp{1.0cm}cccc}
\toprule
$G$ [MPa] & $K$ [MPa] & $E_F$ [MPa] && $G_F$ [MPa] & $K_F$ [MPa] & $G_M$ [MPa] & $K_M$ [MPa]\\
\midrule
1 & 1 & 1 && 50 & 108 & 2 & 10 \\
\bottomrule 
\end{tabular}
\end{table}

\subsection{Comparison of $\psi_1$ and $\psi_2$}

\begin{figure}[htb]
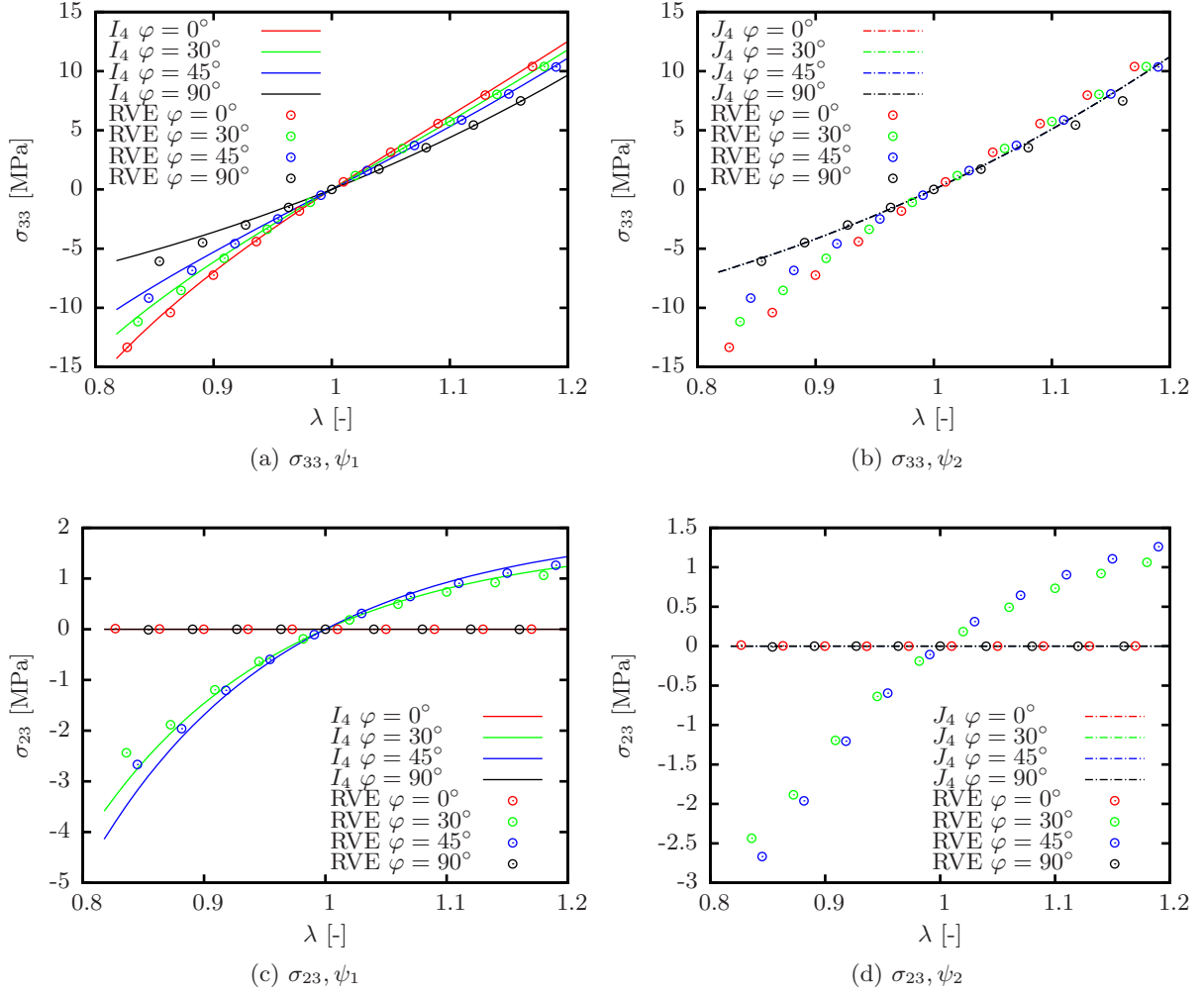

\centering
\subfloat[$\sigma_{33}, \psi_1$\label{sub:dil_normal_I4}]{\footnotesize{\import{identification/}{pic_I4_Dilatation_S33}}}
\subfloat[$\sigma_{33}, \psi_2$\label{sub:dil_normal_J4}]{\footnotesize{\import{identification/}{pic_J4_Dilatation_S33}}}

\subfloat[$\sigma_{23}, \psi_1$\label{sub:dil_shear_I4}]{\footnotesize{\import{identification/}{pic_I4_Dilatation_S23}}}
\subfloat[$\sigma_{23}, \psi_2$\label{sub:dil_shear_J4}]{\footnotesize{\import{identification/}{pic_J4_Dilatation_S23}}}
\caption[]{Comparison of material model's behaviour with data from the RVE (25\% fibres) obtained during purely volumetric deformation. The diagrams show the longitudinal stress as well as the shear stress over the volume stretch for two different energy densities.}
\label{fig:vergleich_I4J4}
\end{figure}

The energy densities $\psi_1$ and $\psi_2$ are fitted to the data from purely volumetric deformation as this points out the differences between both energy densities. \tabref{tab:I4J4} shows the identified material parameters. It can be seen that the shear modulus of both energy densities cannot be determined as the residuum's derivative with respect to the shear modulus equals zero
\begin{equation}
\frac{\partial \Vek{r}}{\partial G} = 0 \qquad,
\label{eq:drdG}
\end{equation}
with $\Vek{r}$ being the residuum. This is expected due to the hydrostatic form of the corresponding deformation gradient $\Ten2F$. The bulk moduli are similar with only 0.8 [MPa] deviation. The main difference can be observed at the fibre's Young's modulus. Whereas $E_F$ of $\psi_2$ has no influence, the value of $E_F$ for the first energy density is changed significantly. The reason behind this lies in the invariant $J_4$. $\Ceins$ always equals $\Ein$ when the deformation gradient has a hydrostatic form. In such case,
\begin{equation}
J_4 = 1 \qquad,\qquad \psi_\mathrm{aniso,2} = 0 \qquad.
\end{equation}
In other words, the anisotropic part of $\psi_2$ vanishes independently from the magnitude of the volume change. \figref{fig:vergleich_I4J4} reveals the physical meaning of this characteristic. \figref{sub:dil_normal_I4} as well as \figref{sub:dil_shear_I4} show that $\psi_1$ catches the anisotropic influence of the fibres. There is a good agreement for $\lambda > 1$ and slight deviations for $\lambda < 1$. \figref{sub:dil_normal_J4} and \figref{sub:dil_shear_J4}, however, demonstrate the energy density's independence from the fibre angle as the curves of the single fibre angles are all identical and there are no shear stresses. Here, only the isotropic part is fitted to the data, meaning the resulting material parameters accord to the best compromise between all the fibre angles. Note that the same behaviour can be observed for almost incompressible fiber and matrix materials.

This comparison points out that \textit{the invariant $J_4$ is not suited to represent the anisotropy in case of a purely volumetric deformations}. That is why only the invariant $I_4$ is used in the following.
\begin{table}[htb]
\centering
\caption[]{Identified material parameters after fitting to the data from purely volumetric deformation using both $I_4$ and $J_4$.}
\label{tab:I4J4}
\begin{tabular}{cccc}
\toprule
$\psi$ & $G$ [MPa] & $K$ [MPa] & $E_F$ [MPa] \\
\midrule
$\psi_1$ & - & 13.2759 & 24.5004 \\
$\psi_2$ & - & 15.4082 & - \\
\bottomrule 
\end{tabular}
\end{table}

\subsection{Minimal set of experiments}

\begin{figure}[ht]
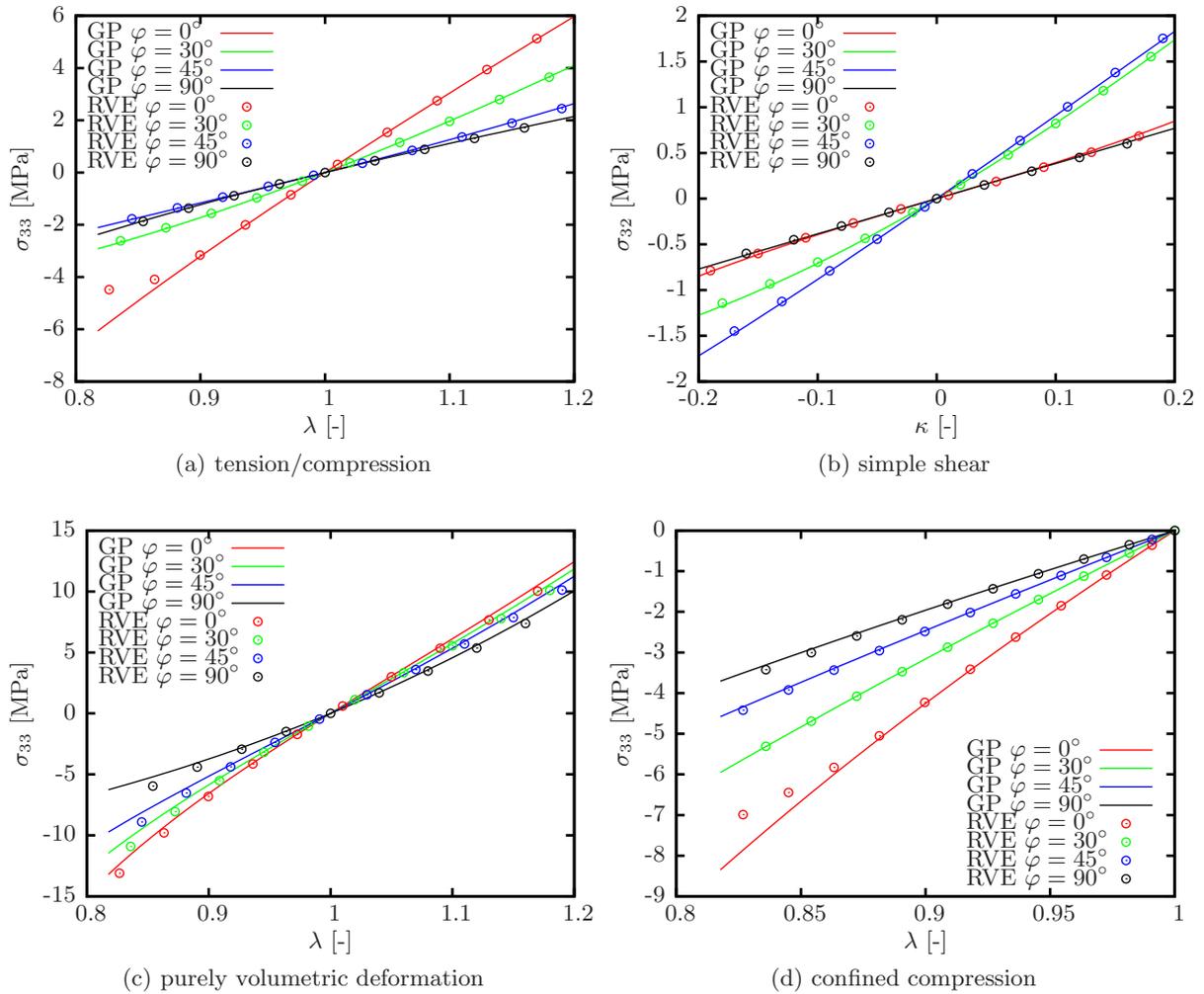

\centering
\subfloat[tension/compression \label{sub:I4_min_zug}]{\footnotesize{\import{identification/}{pic_I4_min_Zug}}}
\subfloat[simple shear \label{sub:I4_min_scherung}]{\footnotesize{\import{identification/}{pic_I4_min_Scherung}}}

\subfloat[purely volumetric deformation \label{sub:I4_min_dilatation}]{\footnotesize{\import{identification/}{pic_I4_min_Dilatation_S33}}}
\subfloat[confined compression \label{sub:I4_min_compression}]{\footnotesize{\import{identification/}{pic_I4_min_DruckDilatation}}}
\caption[]{Comparison of material model's behaviour (Gauss point, GP) with data from the RVE (25\% fibres). The material parameters are fitted to the minimal set of experiments.}
\label{fig:I4_min}
\end{figure}

The identification of material parameters is an expensive and time consuming process. Naturally, the experimental effort should be reduced to a minimum. That is why in this section a minimal set of data for the identification is presented. Each set of data represents a real experiment. As the comparison of \eqref{eq:psi1} and \eqref{eq:psi2} showed, a purely volumetric deformation is not suited to identify the shear modulus. In contrast, a purely isochoric deformation would not be influenced by the bulk modulus. Thus, at least two simulations are necessary. In order to cover the shear modulus and the fibre's Young's modulus, the tension/compression simulation is chosen. The anisotropy is caught by two different values of the fibre angle, i.e. $0^\circ, 90^\circ$. Those are favourable because they do not need special consideration at the fixation of the specimens in a real experiment, as no shear deformations are expected. The data for stretch values $\lambda < 0.9$ are ignored because the material model is not supposed to consider the RVE's sideways motion. The bulk modulus could be covered by the purely volumetric deformation but the related experiment would be very difficult to realise in practice. That is why a confined compression with the only fibre angle $\varphi = 0^\circ$ is chosen. The minimal set of experiments can be seen in \tabref{tab:I4_min_set}. The identification delivers the values in \tabref{tab:I4_min}. This time the resulting material behaviour is compared with the synthetic data of all four simulations in \figref{fig:I4_min}. For the purpose of clarity, only four different fibre angles are presented. Although, only three sets of data are used for the identification, the material model is capable of representing the RVE's behaviour. The only differences can be seen in \figref{sub:I4_min_zug} and \figref{sub:I4_min_compression} for $\lambda < 0.9$ as well as in \figref{sub:I4_min_dilatation} for $\lambda<1$. The latter indicates effects  of the short fibres which are not covered by the energy density $\psi_1$. However, it shall be emphasised again, that the material parameters are only fitted to two simulations with only two different fibre angles, but this is sufficient to describe the material behaviour for arbitrary loading conditions in general simulations.

\begin{minipage}[t]{.45\textwidth}
\begin{table}[H]
\centering
\setlength\abovecaptionskip{0pt}
\setlength\belowcaptionskip{2pt}
\caption[]{Minimal set of experiments}
\label{tab:I4_min_set}
\begin{tabular}{cc}
\toprule
Simulation & Fibre angle \\
\midrule
tension/compression & $0^\circ, 90^\circ$ \\
confined compression & $0^\circ$\\
\bottomrule 
\end{tabular}
\end{table}
\end{minipage}\hfill
\begin{minipage}[t]{.45\textwidth}
\begin{table}[H]
\centering
\setlength\abovecaptionskip{0pt}
\setlength\belowcaptionskip{3pt}
\caption[]{Identified material parameters using the minimal set of simulations.}
\label{tab:I4_min}
\begin{tabular}{ccc}
\toprule
$G$ [MPa] & $K$ [MPa] & $E_F$ [MPa] \\
\midrule
3.8511 & 13.7987 & 20.5426 \\
\bottomrule 
\end{tabular}
\end{table}
\end{minipage}

\subsection{Convex combination}

\begin{figure}[hbt]
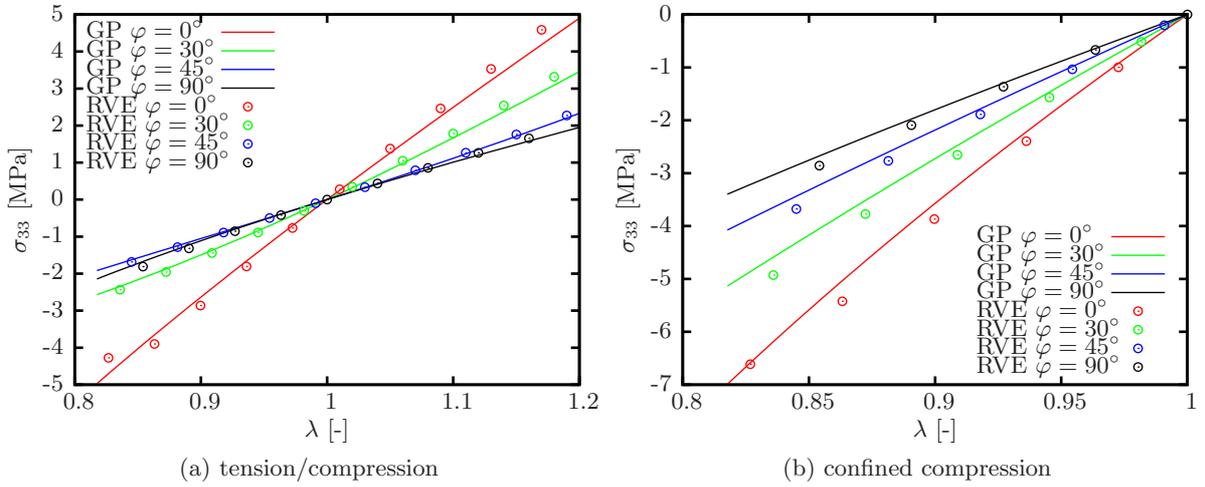

\centering
\subfloat[tension/compression]{\footnotesize{\import{identification/}{pic_f20_min_Zug}}}
\subfloat[confined compression]{\footnotesize{\import{identification/}{pic_f20_min_DruckDilatation}}}
\caption[]{Comparison of material model's behaviour (Gauss point, GP) with data from a RVE containing $20\%$ fibres. The material parameters are obtained from \tabref{tab:convex}.}
\label{fig:convex}
\end{figure}

Energy density $\psi_3$ corresponds to a convex combination with respect to the fibre volume fraction $f_\mathrm{vol}$. \eqref{eq:psi3} allows a direct consideration of the amount of fibres. One interpretation of this ansatz states that one and the same set of parameters $G_M,K_M,G_F,K_F,E_F$ applies to RVEs with the same matrix and fibre materials but different values of $f_\mathrm{vol}$. This approach is tested with the help of three RVEs consisting of $10\%$, $20\%$ and $25\%$ fibres respectively. The identification of the material parameters requires data from RVEs with at least two different values of $f_\mathrm{vol}$. Otherwise, the parameters $G_M$ and $G_F$ as well as $K_M$ and $K_F$ could not be distinguished. Here, the values $f_\mathrm{vol}=10\%$ and $f_\mathrm{vol}=25\%$ are used for the identification. The material parameters are optimised adopting the minimal set of experiments (see \tabref{tab:I4_min_set}). The identified parameters can be found in \tabref{tab:convex}. Using these parameters together with $f_\mathrm{vol}=20\%$, \eqref{eq:psi3} is supposed to predict the behaviour of the RVE which contains 20\% fibres.
\begin{table}[H]
\centering
\caption[]{Identified material parameters using a convex combination of energy densities.}
\label{tab:convex}
\begin{tabular}{cccccc}
\toprule
  & $G_M$ [MPa] & $K_M$ [MPa] & $G_F$ [MPa] & $K_F$ [MPa] & $E_F$ [MPa] \\
\midrule
start value & 1 & 1 & 1 & 1 & 1 \\
identified value &  1.8723 & 9.8631 & 10.0289 & 24.0770 & 79.4441 \\
\bottomrule 
\end{tabular}
\end{table}
\figref{fig:convex} compares the identified material model with the control RVE. The values at the Gauss points reveal a good correspondence to the data obtained by the RVE. Although the short fibres and the matrix material experience different deformations, in contrast to endlessfibre-reinforced materials, the deviations between the RVE data and the predicted values of the identified material model are small. Hence, \eqref{eq:psi3} seems to be suitable for the consideration of variable fibre volume fraction. 


\section{Conclusion}
\label{chap:conclusion}

Three different hyperelastic models to describe transversal isotropic behaviour of short fibre re\-in\-forced materials were discussed. The constitutive equations are designed for large deformations and make use of invariants of the right Cauchy-Green tensor as well as the structural tensor. The material models were fitted to synthetic data obtained with the help of a representative volume element (RVE). The RVE contains unidirectional fibres which are randomly distributed according to several parameters like the fibre volume fraction or the minimal distance between the fibres. The concept of computation models derived from the same base model enables a good comparison of different simulations with the same fibre distribution. Thanks to the refined modelling of the RVE, the effective stress is accurately predicted as long as there are sufficient elements, mainly in longitudinal direction. 

Periodic boundary conditions (PBC) were applied to the RVE as they fulfill the Hill condition per se and do not cause any artificial stiffening or softening. The tensorial formulation of the PBC allows deformations aslant to the fibres. Thus, one and the same model can be used for different fibre angles. Thanks to three auxiliary nodes, the average strain and stress can be directly evaluated. The necessary relations for the implementation of the PBC into \aba were provided.

Four simulations with the RVE revealed its anisotropic properties. The fibre angle has a significant impact on the stiffness at the tension/compression as well as the confined compression. Deformations in the area of big compression lead to an artificial sideways motion of the RVE. The effect of the lengthening and shortening of the fibres during simple shear can be clearly seen. 

The parameters of the three material models were fitted to the synthetic experimental data. It could be pointed out that the mixed invariant for the isochoric right Cauchy-Green tensor ignores purely volumetric deformations. As a consequence, the mixed invariants should be only used for the right Cauchy-Green tensor of the entire deformation. It could be also concluded that a set of only three experiments, i.e. tension/compression at fibre angle $0^\circ, 90^\circ$ and confined compression at fibre angle $0^\circ$, is sufficient to identify material parameters, which cover the whole span of fibre angles as well as several experiments, e.g. simple shear and purely volumetric deformation. This set of experiments was carefully chosen with respect to its practicability. The comparison of the RVEs with different amount of fibres revealed that a convex combination of the energy density according to \eqref{eq:psi3} seems to be capable of interpolating the stress response between different values of $f_\mathrm{vol}$ and a constant set of material parameters. 

The present paper introduced a new method to consider tilted fibres which allows simulations with one and the same RVE under different fibre angles. This method is universally applicable as long as the RVE fulfills the PBC's requirements such as opposing points. In this work, constitutive relations for hyperelastic materials were analysed. However, the use of the \aba \textit{User Subroutine UMAT} enables the implementation of arbitrary material models as long as the required values of stress and stiffness can be provided. Thus, in further research, more complicated effects like viscosity as well as plasticity will be considered.  Furthermore, comparisons of data obtained by the RVE with data from real experiments will be object of research.

\section*{Acknowledgements}

This work was performed within the Federal Cluster of Excellence EXC 1075 \textit{MERGE Technologies for Multifunctional Lightweight Structures} and supported by the German Research Foundation (DFG). Financial support is gratefully acknowledged.


\section*{References}
\bibliographystyle{elsarticle-num}
\bibliography{references}


\end{document}